\documentclass[11pt]{article}
\usepackage{latexsym} \usepackage{epsf}
\usepackage{a4}
\usepackage{amsfonts}
\usepackage{amsmath}
\usepackage{amsthm}
\usepackage{amssymb}
\usepackage{graphics}
\usepackage[cp1251]{inputenc}
\usepackage{bbold}
\usepackage{mathrsfs}
\usepackage{cite}

\renewcommand{\theequation}{\thesection.\arabic{equation}}
\newcounter{subequation}[equation]
\makeatletter

\expandafter\let\expandafter\reset@font\csname reset@font\endcsname

\def\subeqnarray{\arraycolsep1pt
    \def\@eqnnum\stepcounter##1{\stepcounter{subequation}%
        {\reset@font\rm(\theequation\alph{subequation})}}
\jot5mm     \eqnarray}

\makeatother

\def\tr{\mathop{\hbox{\rm tr}}\nolimits}

\def\be{\begin{equation}}
\def\ee{\end{equation}}
\def\bea{\begin{eqnarray}}
\def\eea{\end{eqnarray}}
\def\dd{\partial}
\def\half{\frac{1}{2}}

\def\one#1{#1^{\raise5pt\hbox{$\scriptstyle\!\!\!\!1$}}\,{}}
\def\two#1{#1^{\raise5pt\hbox{$\scriptstyle\!\!\!\!2$}}\,{}}

\def\II{\hbox{{1}\kern-.25em\hbox{l}}}
\makeatletter
\def\binrel@#1{\begingroup
  \setboxz@h{\thinmuskip0mu
    \medmuskip\m@ne mu\thickmuskip\@ne mu
    \setbox\tw@\hbox{$#1\m@th$}\kern-\wd\tw@
    ${}#1{}\m@th$}%
  \edef\@tempa{\endgroup\let\noexpand\binrel@@
    \ifdim\wdz@<\z@ \mathbin
    \else\ifdim\wdz@>\z@ \mathrel
    \else \relax\fi\fi}%
  \@tempa
}
\let\binrel@@\relax
\def\overset#1#2{\binrel@{#2}%
  \binrel@@{\mathop{\kern\z@#2}\limits^{#1}}}
\def\underset#1#2{\binrel@{#2}%
  \binrel@@{\mathop{\kern\z@#2}\limits_{#1}}}
\makeatother
\newfont{\bbd}{msbm10 scaled\magstep1}
\def\C{\hbox{\bbd C}}

\def\NN{\hbox{\bbd N}}
\def\R{\hbox{\bbd R}}
\def\V{\hbox{\bbd V}}
\def\U{\hbox{\bbd U}}

\begin{document}


\begin{titlepage}

\vspace*{1cm}

\begin{center}
{\LARGE \bf{ Baxter operators for arbitrary spin II}}

\vspace{1cm}

{\large \sf D. Chicherin$^{da}$\footnote{\sc e-mail:chicherin@pdmi.ras.ru},
  S. Derkachov$^{a}$\footnote{\sc e-mail:derkach@pdmi.ras.ru}, D.
Karakhanyan$^b$\footnote{\sc e-mail: karakhan@lx2.yerphi.am},
R. Kirschner$^c$\footnote{\sc e-mail:Roland.Kirschner@itp.uni-leipzig.de} \\
}

\vspace{0.5cm}

\begin{itemize}
\item[$^a$]
{\it St. Petersburg Department of Steklov Mathematical Institute
of Russian Academy of Sciences,
Fontanka 27, 191023 St. Petersburg, Russia}
\item[$^b$]
{\it Yerevan Physics Institute, \\
Br. Alikhanian st. 2, 375036 Yerevan, Armenia}
\item[$^c$]
{\it Institut f\"ur Theoretische
Physik, Universit\"at Leipzig, \\
PF 100 920, D-04009 Leipzig, Germany}
\item[$^d$]
{\it Chebyshev Laboratory, St.-Petersburg State University,\\
14th Line, 29b, Saint-Petersburg, 199178 Russia}
\end{itemize}
\end{center}
\vspace{0.5cm}
\begin{abstract}
This paper presents the second part of our study devoted to the
construction of Baxter operators for  the homogeneous closed XXX
spin chain with the quantum space carrying infinite or finite-dimensional
$s\ell_2$ representations. We consider the Baxter operators used
in \cite{BLZ,Shortcut}, formulate their construction uniformly
with the construction of our previous paper. The building blocks
of all global chain operators are derived from the general Yang-Baxter
operators and all operator relations are derived from general
Yang-Baxter relations. This leads naturally to the comparison of both
constructions and allows to connect closely the treatment of
the cases of infinite-dimensional representation of generic spin and
finite-dimensional representations of integer or half-integer spin.
We prove not only the relations between the operators but present also
their explicit forms and expressions for their action on polynomials
representing the quantum states.
\end{abstract}

\vspace{4cm}

\end{titlepage}

\vspace{4cm}

\newpage

{\small \tableofcontents}
\renewcommand{\refname}{References.}
\renewcommand{\thefootnote}{\arabic{footnote}}
\setcounter{footnote}{0} \setcounter{equation}{0}

\newpage

\section{Introduction }
\setcounter{equation}{0}

In our  previous paper~\cite{I}\footnote{In the following \cite {I}
will be referred to as
part I. }
 we have described in detail an approach to the construction
of $\mathrm{Q}$-operators for the
spin chain with the symmetry algebra $s\ell_2$. There exists
another approach to the same problem~\cite{BLZ,Shortcut}.
The aim of the present paper is to unify both approaches in the framework of
the Quantum Inverse Scattering Method (QISM)
~\cite{FST,TTF,KuSk1,Fad} and to establish explicit interrelations.
We choose the homogeneous closed  XXX spin chain as the basic example in
particular because the spin $\half$ case of such a chain has been considered
in~\cite{Shortcut}.

The general strategy is the same in both approaches to Baxter $\mathrm{Q}$-operators
and it appears universal  in the framework of QISM.
Formulated for the finite-dimensional $s\ell_2$ representation case,
 the two main steps of the construction are:
\begin{itemize}
    \item Construction of the general transfer matrix, here for
    spin of quantum spaces  $\ell = \frac{n}{2}, n\in \NN $,
 spin of auxiliary space $s\in \C$
    and regularization parameter $q, |q|<1$ as
\be \label{Tfin}
\mathbf{T}_{s}(u) = \tr_0  \, q^{z_0 \dd_0} \,
\mathbf{R}_{10}\left(u|{\textstyle\frac{n}{2}}, s\right)
\mathbf{R}_{20}\left(u|{\textstyle\frac{n}{2}}, s\right) \cdots
\mathbf{R}_{N0}\left(u|{\textstyle\frac{n}{2}}, s\right) \,. \ee
where $\mathbf{R}_{k0}\left(u|{\textstyle\frac{n}{2}}, s\right)$
is the general solution of the Yang-Baxter equation acting in the
tensor product $\V_n\otimes \U_{-s}$, where $\V_n$ is
$(n+1)$-dimensional and $\U_{-s}$ is infinite-dimensional
irreducible $s\ell_2$-modules. In the case of half-integer spin
$\ell = \frac{n}{2}$ the trace over infinite-dimensional space
diverges so that we need the regularization depending on $q$.
    \item Proof of the following factorization
\be \label{T->QAQBfin} \mathbf{S}\,\mathbf{T}_{s} (u) =
\mathbf{Q}_{B} (u - s) \, \mathbf{Q}_{A} (u + s + 1) =
\mathbf{Q}_{A} (u + s + 1)\, \mathbf{Q}_{B} (u - s)
\,,
\ee
where commuting operators $\mathbf{Q}_{A}$ and $\mathbf{Q}_{B}$
are some transfer matrices
$$
\mathbf{Q}_A(u) = \tr_{0} \, q^{z_0 \dd_0} \, \mathbf{R}^A_{10}(u
)\cdots \mathbf{R}^A_{N0}(u)\ \ \ ;\ \ \ \mathbf{Q}_B(u) = \tr_{0}
\, q^{z_0 \dd_0} \, \mathbf{R}^B_{10}(u)\cdots
\mathbf{R}^B_{N0}(u)\,,
$$
constructed from the local operators $\mathbf{R}^A_{k0}(u)$ and
$\mathbf{R}^B_{k0}(u)$. These local operators are closely related to the
general Yang-Baxter operator
$\mathbf{R}_{k0}\left(u|{\textstyle\frac{n}{2}}, s\right)$
and can be seen as building blocks of the latter because of factorization.
 The
operator $\mathbf{S}$ does not depend on spectral parameter $u$
and commutes with $\mathbf{T}_{s} (u)$ and $\mathbf{Q}_{A}$,
$\mathbf{Q}_{B}$.

\end{itemize}
The factorization~(\ref{T->QAQBfin}) immediately leads to
the determinant representation for the transfer matrix
$\mathrm{t}_m(u)$ constructed with $(m+1)$-dimensional auxiliary space.
The determinant representation is the origin of the  Baxter equation for
operators $\mathbf{Q}_{A}$ and $\mathbf{Q}_{B}$ and of various fusion
relations.
We use the general labels $A$ and $B$, $A,B = +,-$ or $A,B= 1,2$
  because there
are two different pairs of $\mathrm{Q}$-operators,
 $\mathbf{Q}_{1}$
and $\mathbf{Q}_{2}$ studied in part I and $\mathbf{Q}_{+}$ and
$\mathbf{Q}_{-}$ studied in~\cite{Shortcut}.

We have formulated the two basic steps for the case of integer or
half-integer spin at the chain sites, $\ell = \frac{n}{2}$, because this is
the case on which the studies in~\cite{BLZ,Shortcut} are focussed.
The steps are analogous in the case of generic spin $\ell \in \C $.

In part I we have treated first the generic spin case and after this
we have considered the case of integer values of $2\ell$ in analogy.
We have been able to give the relations between the global chain operators,
the general transfer matrix, the Baxter $\mathrm{Q}$-operators
in particular, for the generic and the half-integer spin cases by
explicit expressions.

In the present paper we shall formulate the operators appearing
in~\cite{BLZ,Shortcut}  in such a way that their treatment becomes
unified with the one of part I. The main goal of this paper is the explicit
description of the relations between the two pairs of
$\mathrm{Q}$-operators.

Unlike~\cite{Shortcut}  we prefer to consider  the generic spin case
first and to study from this viewpoint the integer case $2\ell = n$ as a
specific case.
Recall that we denote by $\mathrm{L}$-matrix an operator
acting the tensor product of $s\ell_2$ representation spaces
where one factor of them
is just the fundamental (spin $\half$) representation, whereas the general
Yang-Baxter  $\mathrm{R}$ operator acts on the tensor product with both
factors carrying generic representations.
We use the latter as the most general local chain operator, whereas the
 $\mathrm{L}$-matrix appears as the main building block
in~\cite{Shortcut}. Although the general $\mathrm{R}$ operator is
derived from the $\mathrm{L}$-matrix (by solution of the $RLL$ relation)
we have demonstrated in part I the advantages of the construction relying
on the former.

The presentation is organized as follows. We start from
consideration of the simplest example of the spin one half,
treated in~\cite{Shortcut}. Therefore  in section 2 we
specify  our general formulae for
arbitrary $\ell$ from part I to the case $\ell=\frac{1}{2}$
and give some sketch of the
construction of $\mathbf{Q}_{\pm}$-operators of~\cite{Shortcut}.
By expressing relations in maximally explicit form we
illustrate the relations between both constructions before
addressing the systematic treatment of the general case.

Section 3 is devoted to local objects and their relations. It
is well known~\cite{Backlund,DST1,DST2} that matrices
$\mathrm{L}^{\pm}$ serving as local building blocks for the
construction of $\mathbf{Q}_{\pm}$-operators can be obtained as
special limits of the standard $\mathrm{L}$-operator. Using the
same limits in the case of general $\mathrm{R}$-matrix one obtains
operators $\mathrm{R}^{\pm}$, which are  the
 natural building blocks for construction of
$\mathrm{Q}_{\pm}$-operators if one intends to lift
the whole construction to the case of generic spin $\ell$. We
establish the explicit connection between the two sets of building blocks,
the  operators $\mathrm{R}^1$ and $\mathrm{R}^2$ from part {I} and the
 operators~$\mathrm{R}^{\pm}$ appearing by lifting $\mathrm{L}^{\pm}$.

Next we follow step by step  the same way as in part I.
 The key local relations needed for the factorization of
transfer matrix and commutativity are obtained by applying the mentioned
limiting procedure  to appropriate relations of part I.
We consider the generic spin case in section 4 and then turn to the
case of integer and half-integer spin in section 5.
In particular the action
of the operators on polynomials is written explicitly.

Having derived all necessary formulae for both pairs of
$\mathrm{Q}$-operators it is easy to compare them and to discuss
their advantages and drawbacks.

\section{Spin $\frac{1}{2}$  chain}
\setcounter{equation}{0}

In this section  we specify to the simplest example of
two-dimensional representations at each site of the spin chain.
In this case the general operator
$\mathbf{R}\left(u|{\textstyle\frac{1}{2}}, s\right)$ acting in
the tensor product $\C^2\otimes \U_{-s}$  is simple and
coincides up to normalization and a shift of the spectral
parameter with the $\mathrm{L}$-operator
\begin{equation} \label{R_1/2}
\mathbf{R}\left(u|{\textstyle\frac{1}{2}},s\right) =
-\frac{\Gamma\left(-s-\frac{1}{2}+u\right)}
{\Gamma\left(-s+\frac{1}{2}-u\right)}\cdot
\mathrm{L}\left(u+{\textstyle\frac{1}{2}}\right)\ \ \ ; \ \ \
\mathrm{L}(u) = u + \vec\sigma \otimes \vec{\mathrm{S}} = \left(%
\begin{array}{cc}
  u+\mathrm{S} & \mathrm{S}_{-} \\
  \mathrm{S}_{+} & u-\mathrm{S} \\
\end{array}%
\right)\,,
\end{equation}
 $\vec{\mathrm{S}}$ are generators of $s\ell_2$,
$\mathrm{S}_{\pm} = S_1\pm i S_2$ and $\mathrm{S} = S_3$, in the
infinite-dimensional representation $\U_{-s}$ labeled by spin
$s\in\C$. The usual model for this representation is the space of
polynomials $\C[z]$ where the generators are realized as first order differential
operators
 \be \label{generators} \mathrm{S} =
z\dd -s\ ,\ \mathrm{S}_{-} = -\dd \ ,\ \mathrm{S}_{+} = z^2\dd -
2s\, z \,.\ee

Allowing for a change of normalization and a shift of the spectral parameter,
we take
$\mathrm{L}(u)$ as the building block for the construction of
the transfer matrix
\begin{equation}\label{transfL1/2}
\mathbf{T}_{s}(u|q) = \tr_{z}  \, q^{z \dd} \, \left( u +
\vec\sigma_1\otimes\vec{S}\right)\left( u +
\vec\sigma_2\otimes\vec{S}\right)\cdots\left( u +
\vec\sigma_N\otimes\vec{S}\right)\,.
\end{equation}
Here we use explicit notations for visualization: the quantum space of the
chain is the tensor product of two-dimensional spaces
$\C^2\otimes\cdots\otimes\C^2$ and the $\sigma$-matrices
$\vec\sigma_k$ act in $\C^2$ at  site $k$. The  operators
$\vec{S}$ act in the auxiliary space $\C[z]$ where also the trace $\tr_{z}$
is calculated.

In this section we outline the construction and basic properties of
$\mathbf{Q}$ operators for the example of the spin $\half$ chain.
We start with a sketch of the construction of the
paper~\cite{Shortcut}.    Further we  present
shortly the specification to this example of our construction of part I
and  give a first discussion of the connection between both constructions.

There exists the
useful factorized representation of the operator $\mathrm{L}(u)$
\begin{equation}
\mathrm{L}(u)=\mathrm{L}(u_{1},u_{2}) = \left(%
\begin{array}{cc}
  1 & 0 \\
  z & 1 \\
\end{array}%
\right)\left(%
\begin{array}{cc}
  u_1 & -\partial \\
  0 & u_2 \\
\end{array}%
\right)\left(%
\begin{array}{cc}
  1 & 0 \\
  -z & 1 \\
\end{array}%
\right)\, , \; u_{1}\equiv u-s-1
   \ ,u_{2}\equiv u+s .
\label{LaxFact}
\end{equation}
We have introduced the parameters $u_1$ and $u_2$ along with $u$
and $s$ because they are very convenient for our purposes.
$\mathrm{L}$-operator respects the Yang-Baxter equation in the space
$\C^2 \otimes \C^2 \otimes \U_{-s}$ with Yang's
$\mathrm{R}$-matrix,
\be \mathrm{R}_{ij,nm}(u-v) \cdot
\mathrm{L}_{nr}(u)\cdot \mathrm{L}_{mp}(v) =
\mathrm{L}_{ir}(v)\cdot \mathrm{L}_{jp}(u)\cdot
\mathrm{R}_{rp,nm}(u-v) \label{FCR}, \ee
where $i,j, \cdots = 1,2$
and $ \mathrm{R}_{ij,nm}(u) = u
\cdot\delta_{in}\,\delta_{jm}+\delta_{im}\,\delta_{jn}. $ Besides
of the operator (\ref{LaxFact}) it is also useful to consider  simpler
operators~\cite{Shortcut,DST1,DST2}
\bea \mathrm{L}^{+}(u)  =
u\,\mathrm{e}^{+} + \mathbf{e}\otimes\mathbf{A}^{+} = &
\left(\begin{array}{cc}
u + \dd z & -\dd \\
-z & 1\end{array}\right)  & = \left(\begin{array}{cc}
1 & -\dd \\
0 & 1 \end{array}\right) \cdot
 \left(\begin{array}{cc}
u  & 0 \\
-z & 1 \end{array}\right)\,, \label{L+}
\\
\mathrm{L}^{-}(u)  = u\,\mathrm{e}^{-} +
\mathbf{e}\otimes\mathbf{A}^{-} = & \left(\begin{array}{cc}
1 & -\dd \\
z & u-z \dd
\end{array}\right)  & =
\left(\begin{array}{cc}
1 & 0 \\
z & u \end{array}\right) \cdot
 \left(\begin{array}{cc}
1  & -\dd \\
0 & 1 \end{array}\right)\,. \label{L-} \eea
We use explicit
notations $ \mathrm{e}^{+} = \mathrm{e}_{11}$, $\mathrm{e}^{-} =
\mathrm{e}_{22}$ where $\mathrm{e}_{ik}$ is the standard basis in
the space of two-dimensional matrices and $
\mathbf{e}\otimes\mathbf{A} =
\mathrm{e}_{11}\,\mathrm{A}_{11}+\mathrm{e}_{12}\,\mathrm{A}_{12}+
\mathrm{e}_{21}\,\mathrm{A}_{21}+\mathrm{e}_{22}\,\mathrm{A}_{22}
= \left(\begin{array}{cc}
\mathrm{A}_{11} & \mathrm{A}_{12} \\
\mathrm{A}_{21} & \mathrm{A}_{22}\end{array}\right) $.
We did this to exhibit clearly the separation of the action in quantum space
by matrices from the action in auxiliary space by operators acting on
polynomials in $z$. The matrix elements $\mathrm{A}^{\pm}_{ik}$
are operators on $\U_{-s}$ like $\mathrm{S}_{\pm} $ and $\mathrm{S}
$.  In analogy with ~(\ref{transfL1/2}) we consider the transfer matrices
\be
\mathbf{Q}_{\pm}(u) =
\tr_{z}\, q^{z \dd} \, \left(u\,\mathrm{e}^{\pm}_1 +
\mathbf{e}_1\otimes\mathbf{A}^{\pm}\right)\left(u\,\mathrm{e}^{\pm}_2
+ \mathbf{e}_2\otimes\mathbf{A}^{\pm}\right)\cdots
\left(u\,\mathrm{e}^{\pm}_N +
\mathbf{e}_N\otimes\mathbf{A}^{\pm}\right)\, . \ee

The operators $\mathrm{L}^{\pm}$ also satisfy the Yangian relation
(\ref{FCR}) as well as almost trivial operator $\begin{pmatrix}
1 & 0 \\
z & 1
\end{pmatrix} $.
The Yangian algebra possesses the  important property of
co-multiplication. Let $\mathrm{L}_1(u)$ and $\mathrm{L}_2(u)$ be
two solutions of~(\ref{FCR}), acting in the spaces $\C[z_1]$ and
$\C[z_2]$ correspondingly. Then the matrix
$\mathrm{L}_1(u+\delta_1)\,\mathrm{L}_2(u+\delta_2)$ is also a
solution of~(\ref{FCR}) acting in the space $\C[z_1,z_2] =
\C[z_2]\otimes\C[z_2]$. The possibility to multiply solutions
provides an opportunity to construct many solutions starting from
 simpler ones. In particular it is possible to construct two
more complex solutions of (\ref{FCR}) out of the  above
simpler ones: $\mathrm{L}^{-}_{1}(u_2) \,
\mathrm{L}^{+}_{2}(u_1)$ and $\begin{pmatrix}
1 & 0 \\
z_1 & 1
\end{pmatrix} \, \mathrm{L}_{2}(u_1,u_2)$.
As was pointed out to us by V.Tarasov
 the general theory developed in the papers~\cite{Tar85} implies
the  existence of an intertwining operator for such two
representations of the Yangian algebra.
In this particular case the intertwining operator $e^{z_{2}
\partial_{1}}$ can be exhibited explicitly
\be
\label{trick1/2} e^{z_{2}
\partial_{1}} \cdot \mathrm{L}^{-}_{1}(u_2) \,
\mathrm{L}^{+}_{2}(u_1) =
\begin{pmatrix}
1 & 0 \\
z_1 & 1
\end{pmatrix} \, \mathrm{L}_{2}(u_1,u_2)\cdot e^{z_{2} \dd_1 }\,.
\ee This can be checked by
$$
e^{z_{2} \partial_{1}} \left(\begin{array}{cc}
1 & 0 \\
z_{1} & u_2 \end{array}\right) \left(\begin{array}{cc}
1  & -\dd_{1} \\
0 & 1 \end{array}\right) \cdot \left(\begin{array}{cc}
1 & -\dd_{2} \\
0 & 1 \end{array}\right) \left(\begin{array}{cc}
u_1  & 0 \\
-z_{2} & 1 \end{array}\right) e^{-z_{2} \dd_{1} }=
$$
$$
=
\begin{pmatrix}
1 & 0 \\
z_1 & 1
\end{pmatrix} \cdot
\left(%
\begin{array}{cc}
  1 & 0 \\
  z_{2} & u_2 \\
\end{array}%
\right)\left(%
\begin{array}{cc}
  1 & -\partial_{2} \\
  0 & 1 \\
\end{array}%
\right)\left(%
\begin{array}{cc}
  u_1 & 0 \\
  -z_{2} & 1 \\
\end{array}%
\right)\,
$$
performing similarity transformation in the left hand side for
each matrix factor separately. This local factorization relation
(\ref{trick1/2}) is a cornerstone of the construction.

Similarly one can consider another pair of solutions:
$\mathrm{L}^{+}_{1}(u_1) \, \mathrm{L}^{-}_{2}(u_2)$ and
$\begin{pmatrix}
1 & -\dd_1 \\
0 & 1
\end{pmatrix}
\mathrm{L}_{2}(u_1,u_2)$. Again, these representations of
Yangian algebra must be equivalent  which implies the
existence of the intertwining operator $\mathrm{r}$.
 \be \label{trick'1/2}
\mathrm{r} \cdot
\mathrm{L}^{+}_{1}(u_1) \, \mathrm{L}^{-}_{2}(u_2) =
\begin{pmatrix}
1 & -\dd_1 \\
0 & 1
\end{pmatrix}
\mathrm{L}_{2}(u_1,u_2)\cdot \mathrm{r}\,. \ee
It can be
checked by explicit calculation that the expression for
intertwining operator is
$$
\mathrm{r} = \Gamma(z_2\partial_2+u_1-u_2+1)\,e^{z_{1}
\partial_{2}}.
$$
Following the standard argument of the proof of the commutativity of
the transfer matrix (recalled in part I and applied repeatedly there)
these intertwining relations lead in  the next step to
 corresponding global factorization relations for chain operators.
These relations involve the  transfer matrices
$\mathbf{T}_s, \mathbf{Q}_{\pm}$
acting in the
whole quantum space of the chain.
The first local
relation~(\ref{trick1/2}) leads to
\be
\label{Fact1/2} \frac{1}{1-q}\, \mathbf{T}_{s}(u|q) =
\mathbf{Q}_{-}(u+s) \, \mathbf{Q}_{+}(u-s-1) \ee
while the second
local factorization relation (\ref{trick'1/2}) produces
\be
\label{Fact'1/2} \frac{1}{1-q}\, \mathbf{T}_{s}(u|q) =
\mathbf{Q}_{+}(u-s-1) \, \mathbf{Q}_{-}(u+s) \ee

As one can see the q-regularization is indispensable in order to
ensure  converge of the traces over the infinite dimensional spaces. From
(\ref{Fact1/2}) and (\ref{Fact'1/2}) we deduce commutativity
$$
[\, \mathbf{Q}_{-}(u) \, , \, \mathbf{Q}_{+}(v)\,] = 0
$$
In order to prove the other commutativity relations,
$$
[\, \mathbf{Q}_{-}(u) \, , \, \mathbf{Q}_{-}(v)\,] = [\,
\mathbf{Q}_{+}(u) \, , \, \mathbf{Q}_{+}(v)\,] = 0,
$$
one has to resort to the local intertwining
relations~\cite{DST1,DST2},
$$
\mathrm{P}_{12}\, (1-z_2 \dd_{1})^{u-v} \cdot
\mathrm{L}^{-}_{1}(u) \, \mathrm{L}^{-}_{2}(v) =
\mathrm{L}^{-}_{2}(v)\, \mathrm{L}^{-}_{1}(u) \cdot
\mathrm{P}_{12}\, (1-z_2 \dd_{1})^{u-v},
$$
and
$$
\mathrm{P}_{12} (1+z_{1} \dd_{2})^{u-v} \cdot
\mathrm{L}^{+}_{1}(u) \, \mathrm{L}^{+}_{2}(v) =
\mathrm{L}^{+}_{2}(v)\, \mathrm{L}^{+}_{1}(u) \cdot
\mathrm{P}_{12} (1+z_{1} \dd_{2})^{u-v}.
$$
Thus all operators commute
$$
[\, \mathbf{T}_{s}(u|q) \, , \, \mathbf{Q}_{-}(v)\,] = [\,
\mathbf{T}_{s}(u|q) \, , \, \mathbf{Q}_{+}(v)\,] = 0 .
$$
As the consequence of the factorizations (\ref{Fact1/2}) and
(\ref{Fact'1/2}) one obtains the Baxter equations for operators
$\mathbf{Q}_{\pm}$ \be \mathrm{t}(u|q) \, \mathbf{Q}_{-}(u) = u^N
\, \mathbf{Q}_{-}(u+1) + q (u+1)^N \, \mathbf{Q}_{-}(u-1),
\ee \be
\mathrm{t}(u|q) \, \mathbf{Q}_{+}(u) = q u^N \,
\mathbf{Q}_{+}(u+1) + (u+1)^N \, \mathbf{Q}_{+}(u-1), \ee
where in
the transfer matrix $\mathrm{t}(u|q)$ the trace is taken over
two-dimensional auxiliary space,
$$
\mathrm{t}(u|q) = \tr \begin{pmatrix} q & 0 \\ 0 & 1
\end{pmatrix}
\mathrm{L}_{1}(u)\,\mathrm{L}_{2}(u)\cdots\mathrm{L}_{N}(u).
$$
Explicit formulae for the Baxter operators $\mathbf{Q}_{\pm}$
are presented in  section \ref{explicitfindim}.

Now we specify our formulae from part I for the considered case
$\ell = \frac{1}{2}$ allowing for some change of normalizations
for  simplicity. We start again from (\ref{R_1/2}) relating the
 general operator
$\mathbf{R}\left(u|{\textstyle\frac{1}{2}}, s\right)$ acting in
the tensor product $\C^2\otimes \U_{-s}$  with the
$\mathrm{L}$-operator.
Next we  rewrite the expression for
$\mathbf{R}\left(u-v|{\textstyle\frac{1}{2}}, s\right)$ using
the parametrization
$$
u_1 = u - {\textstyle\frac{3}{2}}, \ u_2 = u +
{\textstyle\frac{1}{2}}\ ;\ v_1 = v - s -1, \ v_2 = v + s\,,
$$
as
$$
\mathbf{R}\left(u_1,u_2|v_1,v_2\right) =
\left(%
\begin{array}{cc}
  1 & 0 \\
  z & 1 \\
\end{array}%
\right)\left(%
\begin{array}{cc}
  u_2-v_2-1 & -\partial \\
  0 & u_1-v_1+1 \\
\end{array}%
\right)\left(%
\begin{array}{cc}
  1 & 0 \\
  -z & 1 \\
\end{array}%
\right)\,,
$$
and consider the limits $v_2\to u_2$ \be
\mathbf{R}(u_1 , u_2|v_1,u_2) = \mathbf{R}^1(u-v_1) = \left(%
\begin{array}{cc}
  1 & 0 \\
  z & 1 \\
\end{array}%
\right)\left(%
\begin{array}{cc}
  -1 & -\partial \\
  0 & u_1-v_1+1 \\
\end{array}%
\right)\left(%
\begin{array}{cc}
  1 & 0 \\
  -z & 1 \\
\end{array}%
\right)\,, \label{R1Fin} \ee and $v_1\to u_1$ \be \label{R2Fin}
\mathbf{R}(u_1 , u_2|u_1,v_2) = \mathbf{R}^2(u-v_2) =
\left(%
\begin{array}{cc}
  1 & 0 \\
  z & 1 \\
\end{array}%
\right)\left(%
\begin{array}{cc}
  u_2-v_2-1 & -\partial \\
  0 & 1 \\
\end{array}%
\right)\left(%
\begin{array}{cc}
  1 & 0 \\
  -z & 1 \\
\end{array}%
\right)\,. \ee
In analogy we consider transfer matrices $\mathbf{Q}_1, \mathbf{Q}_2$
constructed with $\mathbf{R}^1,  \mathbf{R}^2$.
Finally, as a building block for a simple type of transfer matrix denoted by
$\mathbf{S}$ we use the result of the double
limit $v_1\to u_1$ and $v_2\to u_2$
$$
\mathbf{R}(u_1 , u_2|u_1,u_2) = \mathbf{S}_{12} = \left(%
\begin{array}{cc}
  1 & 0 \\
  z & 1 \\
\end{array}%
\right)\left(%
\begin{array}{cc}
  -1 & -\partial \\
  0 & 1 \\
\end{array}%
\right)\left(%
\begin{array}{cc}
  1 & 0 \\
  -z & 1 \\
\end{array}%
\right)\,  .
$$

Our counterpart of the factorization (\ref{Fact1/2}, \ref{Fact'1/2})
looks as follows (compare part I, section 5)
\be \label{T->Q2Q1FinIntr}
\mathbf{S}\,\mathbf{T}_{s} (u) = \mathbf{Q}_{2} (u - s) \,
\mathbf{Q}_{1} (u + s + 1) = \mathbf{Q}_{1} (u + s + 1)\,
\mathbf{Q}_{2} (u - s) \,, \ee where
$$
\mathbf{Q}_1(u) = \tr_{0} \, q^{z_0 \dd_0} \,
\mathbf{R}^1_{10}(u)\cdots
\mathbf{R}^1_{N0}(u)\,,
$$
$$
\mathbf{Q}_2(u) = \tr_{0} \, q^{z_0 \dd_0} \,
\mathbf{R}^2_{10}(u)\cdots
\mathbf{R}^2_{N0}(u)\,,
$$
\be \label{S} \mathbf{S} = \tr_{0} \, q^{z_0\dd_0} \,
\mathbf{S}_{10} \, \mathbf{S}_{20} \cdots \mathbf{S}_{N0} \,.\ee

Of course, the two sets of $\mathrm{Q}$-operators should be
connected. To establish this connection we have  another look at the  definition
of transfer matrix and its factorization to $\mathbf{Q}_{\pm}$-operators,
$$
\mathbf{T}_{s}(u-v) = \tr_{0} q^{z_0 \dd_0}
\mathbf{R}_{10}(u_{1},u_2|v_1,v_{2})\cdots
\mathbf{R}_{N0}(u_{1},u_2|v_1,v_{2}),
$$
$$
\frac{1}{1-q}\, \mathbf{T}_{s}(u-v) = \mathbf{Q}_{-}(u-v+s) \,
\mathbf{Q}_{+}(u-v-s-1).
$$
Specifying  parameters in the first formula we
reduce the transfer matrix $\mathbf{T}_{s}(u-v)$ to the operator
$\mathbf{Q}_{1}$ for $v_2=u_2 \leftrightarrow s = u-v+\frac{1}{2}$
and to the operator $\mathbf{Q}_{2}$ for $v_1=u_1 \leftrightarrow
s = v-u+\frac{1}{2}$. On the other hand the substitution of these
values of parameter $s$ in the factorization formula results in an expression in
terms of $\mathbf{Q}_{\pm}$-operators.
After all one obtains
$$
\mathbf{Q}_{1}(u+1) = (1-q)\,\mathbf{Q}_{+}(-{\textstyle
\frac{3}{2}})\cdot \mathbf{Q}_{-}(u)\ \ ;\ \ \mathbf{Q}_{2}(u+1) =
(1-q)\,\mathbf{Q}_{-}({\textstyle \frac{1}{2}})\cdot
\mathbf{Q}_{+}(u)
$$
The operators $\mathbf{Q}_{+}(-{\textstyle \frac{3}{2}})$ and
$\mathbf{Q}_{-}({\textstyle \frac{1}{2}})$ do not depend on
spectral parameter $u$ and commute with the others. There are
various forms of such interrelations as we discuss below.


\section{Local operators}

\setcounter{equation}{0}

In this section we start the systematic consideration working out the
general construction as outlined in Introduction. We consider first
the operators representing the local building units of the chains.
Actually the $\mathrm{L}$-operator contains the local information
about the system and the $\R$-operators are derived therefrom. On
the other hand  the $\R$-operators are more convenient as building
units acting on the tensor product of quantum and auxiliary spaces
carrying arbitrary representations. The particular case of  spin
$\half$ representation in one of these spaces leads us back to the
$\mathrm{L}$-matrix. The $\R$-operators provide  us with the starting
point from which the different versions of Baxter operators can be
obtained. At first we discuss the operators $\mathrm{L}^{\pm}$
introduced above and then consider their analogues $\R^{\pm}$ on
the level of $\mathrm{R}$-operators.

\subsection{$\mathrm{L}$-operators}
\label{L-operators}

The $s\ell_2$ Lie algebra generators (\ref{generators}) appear in
the evaluation representations of the Yangian algebra, which is
generated by the matrix elements of the $\mathrm{L}$-operator with
the fundamental $RLL$ relation (\ref{FCR}) being the
algebra relations. Besides of the considered standard ones there
are other Yangian representations, described by
$\mathrm{L}$-operators obeying (\ref{FCR}) but different from
(\ref{LaxFact}). These can be obtained from the ordinary
$\mathrm{L}$-operators in the limits $u_2 \to \infty$ or $u_1 \to
\infty$ and will be denoted by $\mathrm{L}^+$ (\ref{L+}) and
$\mathrm{L}^-$ (\ref{L-}) correspondingly. Below we present the
necessary  calculations for completing  the picture. The operators
$\mathrm{L}^{\pm}$ appear in the degenerate self-trapping (DST)
chain model~\cite{DST1,DST2}. On the other hand the ordinary
$\mathrm{L}$-operator can be reconstructed from the degenerate
ones $\mathrm{L}^{\pm}$ by means of the formula (\ref{trick1/2}). For
our purposes it is important that both the reduction and the
reconstruction relations can be lifted from the
$\mathrm{L}$-operators to the $\R$-operator level.

The dependence of the $\mathrm{L}$-operator on  the two parameters
$u_{1},u_{2}$ can be factorized in two ways,
\begin{equation}
\mathrm{L}(u_1,u_2)= \left(\begin{array}{cc}
1 & 0 \\
z& u_2\end{array}\right)\ \left(\begin{array}{cc}
u_1 + \dd z & -\dd \\
-z& 1\end{array}\right)\ = \left(\begin{array}{cc}
1 & -\dd \\
z& u_2-z\dd\end{array}\right)\ \left(\begin{array}{cc}
u_1 & 0 \\
-z& 1\end{array}\right) \label{Lfactor}.
\end{equation}
Let us rewrite this taking into account the notations (\ref{L+}) , (\ref{L-})
\begin{equation}
\mathrm{L}(u_1,u_2)= \left(\begin{array}{cc}
1 & 0 \\
z& u_2\end{array}\right)\cdot\mathrm{L}^{+}(u_1) =
\mathrm{L}^{-}(u_2)\cdot\left(\begin{array}{cc}
u_1 & 0 \\
-z& 1\end{array}\right)\ \
\label{Lfactor'} .
\end{equation}
Using the factorization it is easy to derive what happens with the
$\mathrm{L}$-operator in the limit $u_{2}\rightarrow \infty$
\begin{equation}
\left(\begin{array}{cc}
1 & 0 \\
0 & u_2\end{array}\right)^{-1}\,\mathrm{L}(u_1,u_2)=
\left(\begin{array}{cc}
1 & 0 \\
\frac{z}{u_2}& 1\end{array}\right)\ \cdot \mathrm{L}^{+}(u_1)
\longrightarrow  \mathrm{L}^{+}(u_1)
\end{equation}
and similarly in the limit $u_{1}\rightarrow \infty$
\begin{equation}
\mathrm{L}(u_1,u_2)\left(\begin{array}{cc}
u_1 & 0 \\
0& 1\end{array}\right)^{-1} =
\mathrm{L}^{-}(u_2)\left(\begin{array}{cc}
1 & 0 \\
-\frac{z}{u_1} & 1\end{array}\right) \longrightarrow
\mathrm{L}^{-}(u_2)
\end{equation}
so that asymptotically we have
\begin{equation} \label{asymptL}
\mathrm{L}(u_1,u_2) \overset{u_{1}\to\infty}{\longrightarrow}
\mathrm{L}^{-}(u_2) \cdot \left(\begin{array}{cc}
u_1 & 0 \\
0 & 1\end{array}\right)\ \ ;\ \ \mathrm{L}(u_1,u_2)
\overset{u_{2}\to\infty}{\longrightarrow} \left(\begin{array}{cc}
1 & 0 \\
0 & u_2\end{array}\right)\ \cdot \mathrm{L}^{+}(u_1).
\end{equation}

\subsection{General R-operators}

Let us recall some notations from part I.
The general $\mathrm{R}$-operator acts on the space
$\U_{-\ell_1}\otimes\U_{-\ell_2}$ and is defined
as the $s\ell_2$-invariant solution of the following relation
({\it $\mathrm{RLL}$-relation})
\begin{equation}\label{RLL}
\mathrm{R}_{12}(u_{1},u_2|v_{1},v_2)\,
\mathrm{L}_{1}(u_1,u_2)\,\mathrm{L}_{2}(v_1,v_2)=
\mathrm{L}_{1}(v_1,v_2)\,\mathrm{L}_{2}(u_1,u_2)\,
\mathrm{R}_{12}(u_{1},u_2|v_{1},v_2)
\end{equation}
where \be \label{param} u_1 = u-\ell_1-1 ,\ u_2 = u+\ell_1 ;\ v_1
= v-\ell_2-1 ,\ v_2 = v+\ell_2 . \ee
It is also of interest to define the operators $\mathrm{R}^{1}_{12}$ and
$\mathrm{R}^{2}_{12}$ acting in the space $\U_{-\ell_1}
\otimes \U_{-\ell_2}$  by the following relations
\begin{equation}
\mathrm{R}^{1}_{12}\,
\mathrm{L}_{1}(u_1,u_2)\mathrm{L}_{2}(v_1,v_2)=
\mathrm{L}_{1}(v_1,u_2)\mathrm{L}_{2}(u_1,v_2)\,
\mathrm{R}^{1}_{12}\,, \label{R1}
\end{equation}
\begin{equation}
\mathrm{R}^{2}_{12}\,
\mathrm{L}_{1}(u_1,u_2)\mathrm{L}_{2}(v_1,v_2)=
\mathrm{L}_{1}(u_1,v_2)\mathrm{L}_{2}(v_1,u_2)\,
\mathrm{R}^{2}_{12}\,. \label{R2}
\end{equation}
Upper indices $1,2$ distinguish our two operators and lower
indices as usually shows in what spaces the operators act
nontrivially. In generic situation $\ell_1,\ell_2\in \C$ the solutions
of the defining equations~(\ref{R1}, \ref{R2}) have the form
$$
\mathrm{R}^1_{12}(u_1|v_1,v_2) =
\frac{\Gamma(z_{21}\dd_2+u_1-v_2+1)}{\Gamma(z_{21}\dd_2+v_1-v_2+1)}\,,
$$
\be \label{R1R2} \mathrm{R}^2_{12}(u_1,u_2|v_2) =
\frac{\Gamma(z_{12}\dd_1+u_1-v_2+1)}{\Gamma(z_{12}\dd_1+u_1-u_2+1)}\,.
\ee The general $\mathrm{R}$-operator is factorized as follows
(see part I)
\begin{equation}
\mathrm{R}(u_{1},u_2|v_{1},v_2)=
\mathrm{R}^{1}(u_1|v_1,u_2)\mathrm{R}^{2}(u_1,u_2|v_{2}) =
\mathrm{R}^{2}(v_1,u_2|v_2)\mathrm{R}^{1}(u_1|v_1,v_{2})\,.
\label{Rfact}
\end{equation}
Our purpose is to construct transfer matrices for arbitrary spin in
quantum space. For this we need appropriate local building
blocks. In the previous section we have obtained explicit
expressions for $\mathrm{L}^{\pm}$ by means of specific limiting
procedure. Now we going to present explicit expressions for
operators $\R^{\pm}$, the lift of $\mathrm{L}^{\pm}$ to the
$\mathrm{R}$-operator level. For this purpose we consider the limiting
procedure in $\mathrm{RLL}$-relations and obtain the degenerate
$\mathrm{R}$-operators as the leading asymptotics of
$\mathrm{R}(u_1,u_2|v_1,v_2)$ at large values of one or several of
its parameters $u_1, \, u_2,\, v_1, \, v_2$ like in
(\ref{asymptL}). In this section we do not present the corresponding
calculations being rather simple but technical ones and
collect final results only. All details can be found in Appendix A.
Throughout of this section we assume that the spin parameter in quantum
space is a generic complex number. The case of (half)-integer
quantum spin is considered in  Section \ref{RFinDim}.

\subsubsection{Reduction $\mathrm{R}^1, \mathrm{R}^2 \rightarrow \mathrm{r}^{+}, \mathrm{r}^{-}$}
\label{r+r-}

By taking the limit $v_1 \to \infty$ in the defining equation for the operator
$\mathrm{R}^{1}$ (\ref{R1}) it is not difficult to deduce
\begin{equation}\label{r+LL-}
\mathrm{r}^{+}(u_1|v_{2}) \cdot \mathrm{L}_{1}(u_1,u_2)\,
\mathrm{L}^{-}_{2}(v_2)  = \mathrm{L}^{-}_{1}(u_2)\,
\mathrm{L}_{2}(u_1,v_2)\cdot \mathrm{r}^{+}(u_1|v_{2})
\end{equation}
\begin{equation}\label{r+}
v_1^{z_2\partial_2}\,\mathrm{R}^{1}(u_1|v_1,v_2) \to
\mathrm{r}^{+}(u_1|v_{2}) =
\Gamma(z_{2}\partial_2+u_1-v_2+1)\,e^{z_1\partial_2}
\end{equation}
In analogy with (\ref{asymptL}) we have cancelled out the divergent dilatation
operator from the expression of the operator $\mathrm{R}^1$.
Similarly from (\ref{R2})
one obtains the
reduction $v_2 \to \infty$
\begin{equation}\label{r-LL+}
\mathrm{r}^{-}(u_1|u_2)\cdot \mathrm{L}_{1}(u_1,u_2)\,
\mathrm{L}^{+}_{2}(v_1)  = \mathrm{L}^{+}_{1}(u_1)\,
\mathrm{L}_{2}(v_1,u_2)\cdot \mathrm{r}^{-}(u_1|u_2),
\end{equation}
\begin{equation}\label{r-}
\mathrm{R}^{2}(u_1,u_2|v_2) \,
v^{-z_1 \dd_1}_2 \to \mathrm{r}^{-}(u_1|u_2) = e^{-z_2\partial_1}
\, \frac{(-)^{z_1 \dd_1}}{\Gamma(z_{1}\partial_1+u_1-u_2+1)}
\end{equation}

\subsubsection{Reduction $\mathrm{R} \rightarrow \mathrm{R}^{+}, \mathrm{R}^{-}$}
\label{R+R-}

The reduction in the defining equation of the  $\mathrm{R}$-operator
(\ref{RLL})
at $v_1 \to \infty$ leads to
\begin{equation}\label{R+LL-}
\mathrm{R}^{+}(u_1,u_2|v_{2})\cdot \mathrm{L}_{1}(u_1,u_2)\,
\mathrm{L}^{-}_{2}(v_2)  = \mathrm{L}^{-}_{1}(v_2)\,
\mathrm{L}_{2}(u_1,u_2)\cdot \mathrm{R}^{+}(u_1,u_2|v_{2})
\end{equation}
\begin{equation}\label{R+}
v_1^{z_2\partial_2}\,\mathrm{R}(u_1,u_2|v_1,v_2) \to
\mathrm{R}^{+}(u_1,u_2|v_{2}) = \mathrm{r}^{+}(u_1|u_2)\cdot
\mathrm{R}^{2}(u_1,u_2|v_{2})
\end{equation}
and at $v_2 \to \infty$
\begin{equation}\label{R-LL+}
\mathrm{R}^{-}(u_1,u_2|v_{1})\cdot \mathrm{L}_{1}(u_1,u_2)\,
\mathrm{L}^{+}_{2}(v_1)  = \mathrm{L}^{+}_{1}(v_1)\,
\mathrm{L}_{2}(u_1,u_2)\cdot \mathrm{R}^{-}(u_1,u_2|v_{1})
\end{equation}
\begin{equation}\label{R-}
\mathrm{R}(u_1,u_2|v_1,v_2)\,
v_2^{-z_1\partial_1} \to \mathrm{R}^{-}(u_1,u_2|v_{1}) =
\mathrm{R}^{1}(u_1|v_1,u_2)\cdot \mathrm{r}^{-}(u_1|u_2)
\end{equation}

Consequently we have to our disposal the following factorization formulae:
(\ref{Rfact}) for the general $\R$-operator and (\ref{R+}), (\ref{R-}) for
its two reductions $\R^{+},\,\R^{-}$. In fact factorization
relations for the reduced operators are direct consequence of the factorization
relations for the general $\R$-operator. These formulae have analogues
on the level of transfer matrices as we will see soon.
Now we have the appropriate local building blocks
to be implemented in the  construction of different transfer matrices.

\section{Global objects: commuting transfer matrices}

\label{global} \setcounter{equation}{0}

In the previous section we have introduced the local operators which
concern only one site of the spin chain. Now we turn to the description
of the whole system. We are going to construct various families of
commuting operators -- transfer matrices and
$\mathrm{Q}$-operators.

Let us recall some notations from part I. We distinguish two
versions of the general Yang-Baxter operators acting in the space
$\U_{-\ell_1}\otimes\U_{-\ell_2}$ by the notations
$\mathrm{R}_{12}$ and $\R_{12}$.  The former does not contain the
permutation operator $\mathrm{P}_{12}\,$ whereas the latter does,
so they are related by
$$
\R_{12} = \mathrm{P}_{12}\, \mathrm{R}_{12}.
$$
We will also use the notions $\mathbb{r}^{\pm}$, $\R^{\pm}$, $\R^1$
and $\R^2$ which are related to $\mathrm{r}^{\pm}$,
$\mathrm{R}^{\pm}$, $\mathrm{R}^1$ and $\mathrm{R}^2$ in a similar
way. The general transfer matrix $\mathrm{T}(u)$ for the homogeneous
$\mathrm{XXX}$-spin chain is constructed from the local operators
$\R_{k0}(u|\ell,s)$ acting in the tensor product of quantum space
$\U_{-\ell}$ and auxiliary space $\U_{-s}$. The trace is taken
over the generic infinite-dimensional auxiliary space
\begin{equation} \label{T}
\mathrm{T}_{s}(u|q) = \tr_{0}  \left[  q^{z_0 \dd_0} \, \R_{10} (u|\ell, s)\,\R_{20} (u|\ell,
s) \cdots \R_{N0} (u|\ell, s) \right].
\end{equation}
At fixed spin $\ell$ the  free parameters in the general
transfer-matrix $\mathrm{T}_s(u|q)$ are the spectral parameter $u$
and the spin parameter $s$ in the auxiliary space. We recall the
relation to our four-parameter notation (\ref{param})
$$\R_{k0}(u-v|\ell,s)=  \R_{k0} (u_1,u_2|v_1,v_2), $$
\be \label{param'}
u_1 = u-\ell -1, u_2 = u+\ell, v_1 = v-s-1, v_2 = v+ s .
\ee
In this notation the above definition can be rewritten as
\begin{equation}
\mathrm{T}_{s}(u-v|q) = \tr_{0}  \left[ q^{z_0 \dd_0} \, \R_{10}(u_{1},u_2|v_1,v_{2})\cdots
\R_{N0}(u_{1},u_2|v_1,v_{2}) \right] .
\end{equation}

We assume that the spin in the quantum space $\ell$ is generic. As we have
explained in part I this definition is suited for both finite-dimensional and
infinite-dimensional representations in the quantum space. If the
quantum spin parameter $\ell$ is generic, $\ell\in \mathbb{C}, 2\ell+1 \not \in
\NN $,
the trace in (\ref{T}) is convergent even without
q-regularization. However we shall see shortly that such
regularization is unavoidable in the construction of the operators
$\mathrm{Q}_{\pm}$ for all values of $\ell$.

Our next goal is to obtain the factorization
of the general transfer matrix (\ref{T}) into the product
of two other transfer matrices  built from simpler
local blocks and to prove commutativity properties for
the general transfer matrix and its two factors.
We have carried out this programm in part I
using local the three-term relations.
They serve for our current purposes as well.
Let us quote them here again. At first
we have the general Yang-Baxter equation involving
three general $\R$-operators
\be
\label{rf'} \R_{23}(v_{1},v_2|w_{1},w_2)
\R_{13}(u_{1},u_2|w_1,w_2) \R_{12}(u_1,u_2|v_{1},v_2)=
$$
$$
=\R_{12}(u_{1},u_{2}|v_1,v_2)
\R_{13}(u_1,u_{2}|w_1,w_{2})
\R_{23}(v_{1},v_2|w_{1},w_2) \ee
Then we have the relation involving one operator $\R^1$
and two $\R$-operators
$$
\R^1_{23}(v_1|w_1,w_2)
\R_{13}(u_{1},u_2|w_1,w_2)
\R_{12}(u_{1},u_2|v_1,v_{2}) =
$$
\be \label{rf1'}
=\R_{12}(u_{1},u_2|v_1,w_{2})\R_{13}(u_1,u_{2}|w_1,v_{2})
\R^1_{23}(v_1|w_1,w_2)
\ee
and finally the relation with one operator $\R^2$
and two $\R$-operators
$$
\R^2_{23}(v_1,v_2|w_2) \R_{13}(u_{1},u_2|w_1,w_2)
\R_{12}(u_{1},u_2|v_1,v_{2}) =
$$
\be \label{rf2'} = \R_{12}(u_{1},u_2|w_1,v_{2})
\R_{13}(u_1,u_{2}|v_1,w_{2}) \R^2_{23}(v_{1},v_{2}|w_2). \ee
Starting from these three relations
(\ref{rf'}),(\ref{rf1'}),(\ref{rf2'}) it is possible to derive the
factorization of the general transfer matrix $\mathrm{T}(u)$ into
the product of $\mathrm{Q}$-operators, commutativity of all these
operators and also to obtain the Baxter equation.

\subsection{Factorization and commutativity of the general transfer matrix}

\label{gen_tm}

The general transfer matrix has the remarkable factorization properties
\begin{equation}
\label{T->Q+Q-'} \frac{1}{1-q} \cdot \mathrm{T}_{s} (u-v|q) =
\mathrm{Q}_{+} (u - v_2) \, \mathrm{Q}_{-} (u - v_1) =
\mathrm{Q}_{-} (u - v_1) \, \mathrm{Q}_{+} (u - v_2) \, .
\end{equation}
where the operators $\mathrm{Q}_{+}$ and $\mathrm{Q}_{-}$
are transfer matrices constructed from $\R^{+}$ and $\R^{-}$
\be \label{Q-} \mathrm{Q}_{-}(u- v_1) = \left.\tr_{0} \left[
q^{z_{0} \dd_{0} } \, \R^{-}_{10}(u_1,u_2|v_1)\cdots
\R^{-}_{N0}(u_1,u_2|v_1)\right]\right. \,, \ee
\be \label{Q+}
\mathrm{Q}_{+}(u -v_2) = \left. \tr_{0} \left[q^{z_{0} \dd_{0} }
\, \R^{+}_{10}(u_{1},u_2|v_{2})\cdots
\R^{+}_{N0}(u_{1},u_2|v_{2})\right] \right. \, . \ee
Notice that the dependence on the parameters $v_1$ and $v_2$ results in
a simple shift of the spectral parameter,  $u\to u-v_1$ in the first
operator $\mathrm{Q}_{-}$ and $u\to u-v_2$ in the second one.
Eliminating the redundant shift of the spectral parameter ($u-v \to
u$) we have
\begin{equation}
\label{T->Q+Q-} \frac{1}{1-q} \cdot \mathrm{T}_{s} (u|q) =
\mathrm{Q}_{+} (u - s) \, \mathrm{Q}_{-} (u + s + 1) =
\mathrm{Q}_{-} (u + s + 1) \, \mathrm{Q}_{+} (u - s) \, .
\end{equation}

We derive this  from the
underlying local factorization relations
for building blocks of these general transfer matrices.
Starting from (\ref{rf1'}) and choosing the first space to be the
local quantum space
$\U_{-\ell}$ in site $k$, the second space to be the auxiliary
space $\U_{-s}\sim \C[z_0]$ and the third space to be another copy of the
 auxiliary space $\U_{-s}\sim
\C[z_{0^{\prime}}]$
$$
\R^1_{00^{\prime}}(v_1|w_1,w_2) \,
\R_{k0^{\prime}}(u_{1},u_2|w_1,w_2) \,
\R_{k0}(u_{1},u_2|v_1,v_{2}) =
$$
\be \label{PR1PRPR}
=\R_{k0}(u_{1},u_2|v_1,w_{2}) \,
\R_{k0^{\prime}}(u_1,u_{2}|w_1,v_{2}) \,
\R^1_{00^{\prime}}(v_1|w_1,w_2)
\ee
We have to
do the appropriate limiting procedure in the previous relation
taking at first $w_1 \to \infty$
and then $w_2 \to \infty$ which leads to
$$
\mathrm{P}_{00^{\prime}} \,
(-)^{z_{0^{\prime}}\partial_{0^{\prime}}}\,
e^{z_0\partial_{0^{\prime}}} \cdot
e^{z_{0^{\prime}} \dd_k }
\cdot \R_{k0}(u_{1},u_2|v_1,v_{2}) =
$$
\be \label{localT->-+} = \R^{-}_{k0}(u_{1},u_2|v_1)\cdot
\R_{k0^{\prime}}^{+}(u_1,u_{2}|v_{2})\cdot
\mathrm{P}_{00^{\prime}} \,
(-)^{z_{0^{\prime}}\partial_{0^{\prime}}} \,
e^{z_0\partial_{0^{\prime}}} \ee We present the derivation of
(\ref{localT->-+}) in Appendix B.
This local intertwining relation leads in the standard way to the
relation for the corresponding transfer matrices
$$
\tr_{0^{\prime}} \left[\, \mathrm{q}^{z_{0'} \dd_{0'}}\,
e^{z_{0^{\prime}} \dd_1 } \cdots
 e^{z_{0^{\prime}} \dd_N } \right]
\cdot\tr_{0} \left[\, \mathrm{q}^{z_0 \dd_0}\,
\R_{10}(u_{1},u_2|v_1,v_{2})\cdots
\R_{N0}(u_{1},u_2|v_1,v_{2})\right] =
$$
$$
= \tr_{0} \left[\, \mathrm{q}^{z_0 \dd_0 }\,
\R^{-}_{10}(u_{1},u_2|v_{1})\cdots
\R^{-}_{N0}(u_{1},u_2|v_{1})\right]\cdot \tr_{0^{\prime}} \left[
\, \mathrm{q}^{z_{0'} \dd_{0'}}\,
\R^{+}_{10^{\prime}}(u_1,u_2|v_2)\cdots
\R^{+}_{N0^{\prime}}(u_1,u_2|v_2)\right] \, .
$$
As one can easily see the trace $\tr_{0^{\prime}}
\left[ e^{z_{0^{\prime}} \dd_1 } \cdots
 e^{z_{0^{\prime}} \dd_N } \right]$ does not
 converge so that one needs a regularization. The
regularized expressions contains poles at $\mathrm{q} \to 1$.
Indeed,
we readily compute the trace
$$
\tr_{0^{\prime}}
\left[ \mathrm{q}^{z_{0'} \dd_{0'} } e^{z_{0^{\prime}} \dd_1 } \cdots
e^{z_{0^{\prime}} \dd_N } \right] =
\tr_{0^{\prime}} \mathrm{q}^{z_{0'} \dd_{0'} } = \frac{1}{1-q}
$$
and we obtain the second equality in (\ref{T->Q+Q-}). We see that
the regularization violating $s\ell_2$ symmetry is
unavoidable in the present construction.

Starting from (\ref{rf2'}) instead leads to the first
factorization relation in (\ref{T->Q+Q-}) by analogous steps.
The relation (\ref{T->Q+Q-}) also states the commutativity of transfer
matrices constructed from $\R^+$ and $\R^-$.

Above we have formulated local factorization relations (\ref{R+})
and (\ref{R-}) connecting
$\R^-$ with $\R^1$ and $\R^+$ with $\R^2$:
$$
\R^{-}_{12}(u_1,u_2|v_{1}) =
\mathrm{P}_{12}\,\mathrm{R}^{1}_{12}(u_1|v_1,u_2)\, \mathrm{r}^{-}_{12}(u_1|u_2)
$$
$$
\R^{+}_{12}(u_1,u_2|v_{2}) = \mathrm{P}_{12}\,\mathrm{r}^{+}_{12}(u_1|u_2)\,
\mathrm{R}^{2}_{12}(u_1,u_2|v_{2})
$$
 Analogous relations  hold on the
level of transfer matrices,
\be \label{Q-->Q1q-}
\mathrm{P} \, q^{z_1 \dd_1} \cdot \mathrm{Q}_{-}(u) =
\mathrm{Q}_{1}(u|q) \cdot \mathrm{q}_{-}
\ee
and
\be \label{Q+->q+Q2}
\mathrm{P} \, q^{z_1 \dd_1} \cdot \mathrm{Q}_{+}(u) =
\mathrm{q}_{+} \cdot \mathrm{Q}_{2}(u|q) \, ,
\ee
where operators the $\mathrm{q}_{+}$ and $\mathrm{q}_{-}$ are
auxiliary transfer matrices constructed from $\mathbb{r}^{+}_{k0}$
and  $\mathbb{r}^{-}_{k0}$
\be \label{q+} \mathrm{q}_{+} = \tr_{0}
\left[q^{z_{0} \dd_{0} } \, \mathbb{r}^{+}_{10}(u_1|u_2)\cdots
\mathbb{r}^{+}_{N0}(u_1|u_2)\right] \,, \ee \be \label{q-}
\mathrm{q}_{-} = \tr_{0} \left[q^{z_{0} \dd_{0} } \,
\mathbb{r}^{-}_{10}(u_1|u_2)\cdots
\mathbb{r}^{-}_{N0}(u_1|u_2)\right]\, .\ee They are almost inverse
to each other because choosing the parameters in (\ref{T->Q+Q-}) in a
special way we obtain \be \label{q-q+} \frac{1}{1-q} \cdot
\mathrm{P}\, q^{z_1\dd_1} = \mathrm{q}_{-} \cdot \mathrm{q}_{+} =
\mathrm{q}_{+} \cdot \mathrm{q}_{-} \ee

The operators $\mathrm{Q}_1$ and $\mathrm{Q}_2$ are $q$-regularized
Baxter operators constructed in part I
\be \label{regQ1}
\mathrm{Q}_1(u-v_1|q) = \tr_{0} \left[ \, q^{z_0 \dd_0} \,
\R^{1}_{10}(u_1|v_1,u_{2})\cdots
\R^{1}_{N0}(u_1|v_1,u_{2})\right]\,, \ee \be \label{regQ2}
\mathrm{Q}_2(u-v_2|q) = \tr_{0} \left[ \, q^{z_0 \dd_0} \,
\R^{2}_{10}(u_{1},u_2|v_{2})\cdots
\R^{2}_{N0}(u_{1},u_2|v_{2})\right]\,,\ee which obey the
factorization relation \be \label{regT->Q1Q2} \mathrm{P}\, q^{z_1
\dd_1}\cdot\mathrm{T}_{s} (u|q) = \mathrm{Q}_2 (u - s|q) \,
\mathrm{Q}_1 (u + s + 1|q) = \mathrm{Q}_1 (u + s + 1|q) \,
\mathrm{Q}_2 (u - s|q) \ee and commute among themselves. In
Appendix B we list the local factorization relations which produce
(\ref{Q-->Q1q-}) by implementing the reduction in (\ref{PR1PRPR}).

Thus we have derived all factorization properties of transfer matrices
from underlying local relations which are certain reductions of
(\ref{rf1'}) and (\ref{rf2'}). With respect to the commutativity properties
it holds that $q$-regularized transfer matrices constructed from local blocks
$\R, \R^{\pm}, \mathbb{r}^{\pm}, \R^{1}, \R^{2}$ and $\mathrm{P}_{k0}$ commute
among themselves. We deduce these properties from the underlying local relations
which now are reductions of the general Yang-Baxter relation (\ref{rf'}).
The corresponding calculations can be found in Appendix B.

The transfer matrices $\mathrm{Q}_{+}$ and $\mathrm{Q}_{-}$
constructed from operators $\R^{+}$ and $\R^{-}$ have all
properties of the $\mathrm{Q}$-operators, introduced by
R.Baxter~\cite{Baxter1,Baxter2}.  All commutativity properties
have been proven already \be \label{commQ+Q-} [ \, \mathrm{T}_{s}(u|q)
\, , \, \mathrm{Q}_+ (v) \, ] = [ \, \mathrm{T}_{s}(u|q) \, , \,
\mathrm{Q}_- (v) \, ] = \left[ \, \mathrm{Q}_{+}(u) \, , \,
\mathrm{Q}_{-}(v) \, \right] = 0\, \ee
$$
\left[ \, \mathrm{Q}_{+}(u) \, , \, \mathrm{Q}_{+}(v) \, \right] =
\left[ \, \mathrm{Q}_{-}(u) \, , \, \mathrm{Q}_{-}(v) \, \right] =
0\,
$$
so that we shall focus in the
following on the Baxter equation for $\mathrm{Q}_{\pm}$.

In part I we have shown how to prove the Baxter equations staring from
$\mathrm{RLL}$-relation (\ref{R1}) and (\ref{R2}) (restriction to
invariant two dimensional subspace of the tree-term relations
(\ref{rf1'}) and (\ref{rf2'})). One can accomplish the analogous
construction for the present pair of Baxter operators using the
$\mathrm{RLL}$-relations (\ref{R+LL-}) and (\ref{R-LL+}).

But it is much simpler to use the connection between the operators
$\mathrm{Q}_{+},\,\mathrm{Q}_{-}$ and the operators $\mathrm{Q}_1$,
$\mathrm{Q}_2$ expressed by the factorization relations
(\ref{Q-->Q1q-}) and (\ref{Q+->q+Q2}). We have already mentioned
above that all families of transfer matrices
$\mathrm{Q}_{1,2}(u|q)$, $\mathrm{Q}_{\pm}(u)$,
$\mathrm{q}_{\pm}$, $\mathrm{P} \, q^{z_1 \dd_1}$ commute among
themselves.

The operators $\mathrm{Q}_{1,2}(u|q)$ respect the following Baxter
equation (compare part I)\footnote{
In order to obtain Baxter relation in the standard form we change normalization of
$\mathrm{R}$-operators
$\mathrm{R}(u_1 , u_2|v_1 , v_2) \to (-1)^{u_1-v_1}\mathrm{R}(u_1 , u_2|v_1 , v_2)$ ,
$\mathrm{R}^{-}(u_1,u_2|v_1) \to (-1)^{u_1-v_1} \mathrm{R}^{-}(u_1,u_2|v_1)$ and
$\mathrm{R}^1(u_1|v_1 , v_2) \to (-1)^{u_1-v_1}\mathrm{R}^1(u_1|v_1 , v_2)$.
In the other parts of the paper we do not retain this normalization factor since
it can be restored easily. } 

\be \label{qBaxter1}
\mathrm{t}(u|q)\mathrm{Q}_1(u|q) =\mathrm{Q}_1(u+1|q)+  q\cdot(u_1
u_2)^N \cdot\mathrm{Q}_1(u-1|q)\, \ee

\be \mathrm{t}(u|q) \,
\mathrm{Q}_2(u|q) = q\cdot \mathrm{Q}_2(u+1|q)+ (u_1 u_2)^N
\cdot\mathrm{Q}_2(u-1|q)\, \label{qBaxter2}
\ee
where
\be \label{regt} \mathrm{t}(u|q) = \tr
\begin{pmatrix} q & 0 \\ 0 & 1 \end{pmatrix}
\mathrm{L}_1(u)\mathrm{L}_2(u)\cdots \mathrm{L}_N(u)\,. \ee
Then using the relation between the two sets of Baxter operators and
the commutativity of all transfer matrices we obtain immediately
the desired Baxter equations
\be \label{Baxter-}
\mathrm{t}(u|q)\,\mathrm{Q}_{-}(u) =\mathrm{Q}_{-}(u+1) + q\cdot(u_1
u_2)^N \cdot\mathrm{Q}_{-}(u-1) \,,
\ee
\be \label{Baxter+} \mathrm{t}(u|q) \,
\mathrm{Q}_{+}(u) = q\cdot \mathrm{Q}_{+}(u+1) + (u_1 u_2)^N
\cdot\mathrm{Q}_{+}(u-1) \, ,
\ee
since neither $\mathrm{q}_{\pm}$ nor $\mathrm{P} \, q^{z_1 \dd_1}$
 depend on the spectral parameter.

\subsection{Explicit action on polynomials}

\label{explicit}

The previous construction of Baxter operators is a mostly algebraic one.
Now we are going to derive explicit formulae
which show how the constructed operators act on polynomials.
At first we calculate the traces in the definitions of
the simple operators $\mathrm{q}_{+}$ and $\mathrm{q}_{-}$.
Then we take into account the explicit formulae for $\mathrm{Q}_{1,2}(u|q)$
obtained in part I
and finally combine them by means of the factorization relations
(\ref{Q-->Q1q-}), (\ref{Q+->q+Q2}) .

We transform (\ref{q-}), (\ref{q+}) using (\ref{r-}) and
(\ref{r+}) \be \label{q-factor} \mathrm{q}_{-} = \tr_{0}\left[ \,
q^{z_{0} \dd_{0} } \, \mathrm{P}_{10} \, e^{-z_0 \dd_1} \cdots
\mathrm{P}_{N0} \, e^{-z_0 \dd_N} \right] \cdot \Pi_{-} \ee \be
\label{q+factor} \mathrm{q}_{+} = \Pi_{+} \cdot \tr_{0}\left[ \,
q^{z_{0} \dd_{0} } \, \mathrm{P}_{10}\, e^{z_1 \dd_0} \cdots
\mathrm{P}_{N0} \, e^{z_N \dd_0} \, \right] \ee where we suppress
for brevity the dependence on $\ell$ in our notations and extract
the operators
$$
\Pi_{-} \equiv \prod^{N}_{k = 1}{ \frac{(-)^{z_k
\dd_k}}{\Gamma(z_{k}\partial_k - 2 \ell )} } \qquad , \qquad
\Pi_{+} \equiv \prod^{N}_{k = 1}{\Gamma( z_k \dd_k - 2 \ell)}\,.
$$
These operators are relatives of the projectors on the
finite-dimensional subspace that appear in the case of half-integer $\ell$
used in part I. The trace in (\ref{q-factor}) can be calculated
explicitly (compare part I, section 4.3 and Appendix B)
$$
\tr_{0}\left[ \, q^{z_{0} \dd_{0} } \, \mathrm{P}_{10} \, e^{-z_0
\dd_1} \cdots \mathrm{P}_{N0} \, e^{-z_0 \dd_N} \right] =
\mathrm{P} \, q^{z_1 \dd_1} \left. e^{-z_2 \dd_1} e^{-z_3 \dd_2}
\cdots e^{-z_0 \dd_N} \right|_{z_0 = \frac{z_1}{q}},
$$
and this operator acts on polynomials in the  following way
$$
\tr_{0}\left[ \, q^{z_{0} \dd_{0} } \, \mathrm{P}_{10} \, e^{-z_0
\dd_1} \cdots \mathrm{P}_{N0} \, e^{-z_0 \dd_N} \right] \cdot
\Psi(z_1, \cdots , z_N ) =
$$
\be \label{q-trace} = \Psi( q z_N - z_1, z_1 - z_2 , \cdots ,
z_{N-1}- z_N) \,. \ee

In order to calculate the trace in (\ref{q+factor}) we act on the
function $z_0^k \, \Psi(z_1, \cdots , z_N)$
$$
q^{z_{0} \dd_{0} } \, \mathrm{P}_{10} \, e^{z_1 \dd_0} \cdots
\mathrm{P}_{N0} \, e^{z_N \dd_0} \cdot z_0^k \, \Psi(z_1, \cdots ,
z_N) =
$$
$$
= (q z_0 + z_1 + \cdots + z_N)^k \, \Psi( q z_0 , q z_0 + z_1 , q
z_0 + z_ 1 + z_2 , \cdots , q z_0 + z_1 + \cdots + z_N)
$$
and then apply the formula $ \sum_{k=0}^{\infty}\frac{1}{k!} \dd^k_{0}
\cdot\left(a+ b\cdot z_0\right)^k \Phi(z_0)\biggl|_{z_0=0} =
\frac{1}{1-b}\cdot \Phi\left(\frac{a}{1-b}\right)$ proven
in part I (Appendix B), so that one obtains
$$
\tr_{0}\left[ \, q^{z_{0} \dd_{0} } \, \mathrm{P}_{10}\, e^{z_1
\dd_0} \cdots \mathrm{P}_{N0} \, e^{z_N \dd_0} \, \right] \,
\Psi(z_1, \cdots , z_N) =
$$
\be \label{trq+} =\frac{1}{1-q} \cdot \left.\Psi( z_0 , z_0 + z_1
, z_0 + z_1 + z_2 , \cdots , z_0 + z_1 + \cdots + z_{N-1} )
\right|_{z_0 \to \frac{q}{1-q}\cdot(z_1+ \cdots + z_N)}\,. \ee

Let us recall the  formulae for $\mathrm{Q}_{1,2}(u|q)$ from
part I.
The action of the operator
$\mathrm{Q}_1(u|q)$ on polynomials is
\begin{equation}\label{bfQ1q}
\mathrm{Q}_{1}(u|q)\,\Psi(\vec{z}) =
\left.\mathrm{R}_1(\lambda_1\partial_{\lambda_1})\cdots
\mathrm{R}_1(\lambda_N\partial_{\lambda_N})\right|_{\lambda=1}\cdot
\frac{1}{ 1 - q\bar{\lambda}_1 \cdots \bar{\lambda}_N } \cdot
\Psi(\Lambda_q^{\prime -1}\, \vec{z} \, ) \,,
\end{equation}
where
$$
\Lambda_q^{\prime} = \begin{pmatrix}
1-\frac{1}{\lambda_1} & \frac{1}{\lambda_1} & 0 & 0 & \hdotsfor{2} & 0 \\
0 & 1-\frac{1}{\lambda_2} & \frac{1}{\lambda_2} & 0 & \hdotsfor{2} & 0 \\
0 & 0 & 1-\frac{1}{\lambda_3} & \frac{1}{\lambda_3} & \hdotsfor{2} & 0 \\
\hdotsfor{7}  \\
0 & 0 & \hdotsfor{3} & 1-\frac{1}{\lambda_{N-1}} & \frac{1}{\lambda_{N-1}} \\
\frac{1}{q \lambda_{N}} & 0 & 0 & 0 & \cdots & 0 &
1-\frac{1}{\lambda_{N}}
\end{pmatrix} \, , \,\,  \mathrm{R}_1(x) \equiv \frac{\Gamma\left(x-2\ell\right)}
{\Gamma\left(x+1-\ell-u\right)}\
$$
In contrast to $\mathrm{Q}_1$ the renormalized operator $\mathrm{Q}_2$
possess a more elegant representation
\begin{equation} \label{Q}
\mathrm{Q}(u|q) = \frac{\Gamma^N(-2\ell)}{\Gamma^N(-\ell+u)}\cdot
\mathrm{Q}_2(u|q)\, .
\end{equation}
Indeed, its action on the generating function of the representation
in quantum space looks very simple
\begin{equation}
\mathrm{Q} (u|q): (1-x_{1} z_{1})^{2\ell}\cdots(1-x_{N}
z_{N})^{2\ell} \mapsto \label{regQ24}
\end{equation}
$$
\mapsto (1-q \,x_{1} z_{N})^{\ell-u}(1-x_{1}
z_{1})^{\ell+u}\cdots(1-x_{N} z_{N-1})^{\ell-u}(1-x_{N}
z_{N})^{\ell+u} \, .
$$
Now we combine the factorization relation (\ref{Q-->Q1q-}),
the expression for $\mathrm{q}_{+}$ (\ref{q+factor}), (\ref{trq+}),
expression for $\mathrm{Q}(u|q)$ (\ref{regQ24})
and introduce the renormalized operator
\be \label{Q(+)} \mathrm{Q}^{+}(u) =
\frac{1-q}{\Gamma^N(-\ell+u)}\cdot \mathrm{Q}_{+}(u) \ee
which maps
polynomials in $z_1 \cdots z_N$ to polynomials in $u\,,\,z_1
\cdots z_N$
$$
\mathrm{Q}^{+}(u) \,:\, \C[z_1\cdots z_N] \to
\C[u,z_1 \cdots z_N]
$$
We obtain that the renormalized Baxter operator
 acts on the generating function as follows
\begin{equation}\label{Q(+)pol}
\mathrm{Q}^{+}(u) \, : \, \prod^{N}_{k = 1}(1-x_{k}
z_{k})^{2\ell} \, \mapsto \, \prod^{N}_{k = 1}{\frac{\Gamma( z_k
\dd_k - 2 \ell)}{\Gamma(-2\ell)}} \times
\end{equation}
$$ \times
\prod^{N}_{k = 1} \left( 1 - \frac{x_k}{1-q}\cdot( z_{1,k-1} + q
z_{k,N} ) \right)^{\ell-u} \left( 1 - \frac{x_k}{1-q}\cdot(
z_{1,k} + q z_{k+1,N} ) \right)^{\ell+u}\,,
$$
where
$$
z_{1,k} \equiv z_1 + z_2 + \cdots + z_k \qquad ; \qquad z_{k,N} \equiv z_k
+ z_{k+1} + \cdots + z_N\,.
$$
$\mathrm{Q}^{+}(u)$ is normalized in such a way that
$\mathrm{Q}^{+} (u): 1 \mapsto 1$ .

Combining the factorization relation (\ref{Q-->Q1q-}),
the expression for $\mathrm{q}_{-}$ (\ref{q-factor}), (\ref{q-trace})
and the expression for $\mathrm{Q}_1(u|q)$ (\ref{bfQ1q})
we can calculate explicitly how $\mathrm{Q}_{-}$ acts on polynomials.

\section{Finite-dimensional representations}

\label{finiteDim}

\setcounter{equation}{0}

In the previous section we have assumed the spin parameter $\ell$ to
be a generic complex number and consequently the quantum space of
the model was infinite-dimensional. Under this assumptions we have
concentrated on the algebraic properties of $s\ell_2$-invariant
transfer matrices constructed from the operators $\R^{+}$ and $\R^{-}$
and have demonstrated the key properties that allows to call them
$\mathrm{Q}$-operators: commutativity and Baxter equation. Now we
are going to consider the special situation of integer or half-integer spin
$\ell$. In this case the infinite-dimensional representation
becomes reducible and there appears an invariant subspace which is the
standard finite-dimensional irreducible representation labeled by
(half)-integer spin $\ell$.

It turns out that the previous construction for infinite-dimensional
representations can be transferred straightforwardly to the finite-dimensional
case. We only have to substitute integer or half-integer values of $\ell$
in the
formulae obtained above and restrict the operators to the irreducible
finite-dimensional subspace. In doing so we do not face any obstacles.
In other words, the  Baxter operators $\mathrm{Q}_{\pm}$ and the general transfer
 matrix $\mathrm{T}$ constructed above
do not map beyond the detached finite-dimensional subspace
at (half)-integer $\ell$.

Our presentation will be the following:
\begin{itemize}
\item We shall work with the restriction of
the general $\R$-operator and the operators $\R^{+},\,\R^{-}$
to the invariant subspace appearing for
half-integer or integer values of the spin.

\item We use these restricted operators $\R,\,\R^{+},\,\R^{-}$
as building blocks for
the construction of the corresponding transfer matrices.

\item Using local relations we prove the factorization of the general
transfer matrix into the product of the corresponding
$\mathrm{Q}$-operators.

\end{itemize}

To start with we consider simple examples of restriction
to invariant subspace in the next subsection.

\subsection{Examples of restriction: two-dimensional representations}

\label{R->L}

Consider the general $\R$-operator (\ref{Rfact}). It acts in the space
$\U_{-\ell}\otimes\U_{-s}$
and has the form
\begin{equation} \label{PR}
\R_{12} (u|\ell,s) =
\mathrm{P}_{12}\cdot\frac{\Gamma(z_{21}\dd_2-2\ell)}
{\Gamma(z_{21}\dd_2-\ell-s-u)}
\cdot\frac{\Gamma(z_{12}\dd_1-\ell-s+u)}
{\Gamma(z_{12}\dd_1-2\ell)}\,
\end{equation}
after  passing to the spin  parameter notation (\ref{param'})
\be  u_1 = u-\ell-1 ,\ u_2 = u+\ell ;\ v_1
= v-s-1 ,\ v_2 = v+s \,.
\ee

In discrete points $\ell = \frac{n}{2}, n  = 0,1,2,3\cdots $
the module $\U_{-\ell}$ becomes reducible. Therefore it is
possible to restrict the $\R$-operator to the space
$\V_n\otimes\U_{-s}$.
In the case $\ell = \frac{1}{2}$ we have
$$
\V_{1}\otimes\U_{-s}\sim
\C^2\otimes\U_{-s}
$$
and the restricted $\R$-operator acts on functions of the form
\be \label{fun}
\Psi(z_1,z_2) = \phi(z_2) + z_1 \psi(z_2).
\ee
In part I  (section 5.1) we have performed the detailed
calculations of such a restriction
analyzing the limit $\varepsilon \to 0, \ 2 \ell = n - \varepsilon$
and obtained that in the basis $\mathbf{e}_1 = -z_1 , \mathbf{e}_2 = 1$
restricted $\R$-operator has the form
\begin{equation}
\R
\left(u|{\textstyle\frac{1}{2}},s\right)\left|_{\mathbb{V}_{1}}\right.
= -\frac{\Gamma\left(-s-\frac{1}{2}+u\right)}
{\Gamma\left(-s+\frac{1}{2}-u\right)}\cdot
\mathrm{L}\left(u+{\textstyle\frac{1}{2}}|s\right) \, .
\end{equation}

One can easily perform similar calculations for the operators
$\R^{+}$ and $\R^{-}$ (\ref{R+}), (\ref{R-}). In
 the spin parameter notations they have the form
\be \label{R+expl}
\R^{+}_{12}\left( u|\ell,s \right)
= \mathrm{P}_{12}
\cdot \Gamma(z_{2}\partial_2 - 2 \ell)\,e^{z_1\partial_2} \cdot
\frac{\Gamma(z_{12}\dd_1-\ell-s+u)}{\Gamma(z_{12}\dd_1-2\ell)} \, ,
\ee
\be \label{R-expl}
\R^{-}_{12}\left( u|\ell,s \right) = \mathrm{P}_{12} \cdot
\frac{\Gamma(z_{21}\dd_2 - 2\ell)}{\Gamma(z_{21}\dd_2-\ell-s-u)}
\cdot e^{-z_2\partial_1} \frac{(-)^{z_1
\dd_1}}{\Gamma(z_1\partial_1 - 2 \ell)} \, .
\ee
Using the notations (\ref{L+}), (\ref{L-}) we have
\be \label{R+restr}
\R^{+}\left( u|{\textstyle \frac{1}{2}},s \right)
|_{\mathbb{V}^1} = \Gamma(-s - {\textstyle \frac{1}{2} }+u) \cdot
\begin{pmatrix}
-1 & 0 \\
0 & 1
\end{pmatrix}\cdot
\mathrm{L}^{+}(u-s-\textstyle{\frac{1}{2}}),
\ee
\be \label{R-restr}
\R^{-}\left( u|{\textstyle \frac{1}{2}},s \right)
|_{\mathbb{V}^1} = \Gamma^{-1}(-s + {\textstyle \frac{1}{2} }- u)
\cdot
\mathrm{L}^{-}(u+s+\textstyle{\frac{1}{2}}) \cdot
\begin{pmatrix}
1 & 0 \\
0 & -1
\end{pmatrix}\, .
\ee

This example explicitly shows that contrary to the operator
$\R^{2}$ studied in  part I, the operators $\R^{+},\,\R^{-}$ share the
important property with the general $\R$-operator -- they all preserve
the
invariant two-dimensional subspace for half-integer spin. As a
consequence the properties of the general transfer matrix $\mathrm{T}$
as well as the operators $\mathrm{Q}_{-},\,\mathrm{Q}_{+}$ are also similar
with respect to the restriction to the finite-dimensional subspace.

Now we are going to consider the restriction of
the  simple operator $e^{-z_{1} \dd_2}$.
It acts on the space of functions (\ref{fun})
$$
e^{-z_{1} \dd_2} \, \left[ \phi(z_2) + z_1 \psi(z_2) \right] =
\phi(z_2) - z_2 \psi(z_2) + z_1 \psi(z_2)
$$
In the  basis form we have
$$
e^{-z_{1} \dd_2} \,\mathbf{e}_1 = \mathbf{e}_1 \cdot 1  + \mathbf{e}_2 \cdot z_2
$$
$$
e^{-z_{1} \dd_2} \,\mathbf{e}_2 = \mathbf{e}_2 \cdot 1
$$
and in matrix form
\be \label{expExpl}
e^{-z_{1} \dd_2} |_{\mathbb{V}^1}=
\begin{pmatrix}
1 & 0 \\
z_2 & 1
\end{pmatrix}\,.
\ee

Let us now turn to the local relation (\ref{localT->-+}) which produces the
factorization relation for the general transfer matrix and
transform it to the form (indices $1\,,2$ denote two auxiliary
spaces and index $0$ denotes the quantum space in one site of the
chain) \be e^{z_{2} \partial_{1}} \cdot
\R^{-}_{01}(u_{1},u_2|v_1)\cdot \R_{02}^{+}(u_1,u_{2}|v_{2})=
e^{-z_1\dd_0}\cdot\R_{02}(u_{1},u_2|v_1,v_{2}) \cdot e^{z_{2}
\dd_1}\, \ee Then we perform the restriction of this relation to
the two-dimensional subspace in quantum space at $\ell=\frac{1}{2}$.
We take into account (\ref{R+expl}), (\ref{R-expl}),
(\ref{expExpl}) and obtain (\ref{trick1/2})
\be \label{trick}
e^{z_{2}
\partial_{1}} \cdot \mathrm{L}^{-}_{1}(u_2) \,
\mathrm{L}^{+}_{2}(u_1) =
\begin{pmatrix}
1 & 0 \\
z_1 & 1
\end{pmatrix} \, \mathrm{L}_{2}(u_1,u_2)\cdot e^{z_{2} \dd_1 }\,
\ee
which underlies the construction of~\cite{Shortcut}.

Similarly one
can rewrite (\ref{localT->+-}) in the form
$$
\mathrm{r}^{+}_{12}(v_1|v_2) \cdot \R_{01}^{+}(u_1,u_{2}|v_{2}) \cdot
\R^{-}_{02}(u_{1},u_2|v_1)  =
$$
$$
= \Gamma(z_0 \dd_0 + u_1 - u_2+1) \, e^{z_0 \, \dd_1} \, \Gamma^{-1}(z_0 \dd_0 + u_1 - u_2+1) \cdot
(-)^{z_2\dd_2}\,\R_{02}(u_{1},u_2|v_1,v_{2})\,(-)^{z_2\dd_2}\cdot\mathrm{r}^{+}_{12}(v_1|v_2)
$$
and perform the restriction
$$
\Gamma(z_0 \dd_0 -2\ell) \, e^{z_0 \, \dd_1} \, \Gamma^{-1}(z_0 \dd_0 -2\ell)
\left|_{\mathbb{V}^{1},\, \ell\to\frac{1}{2}}\right. =
\begin{pmatrix}
1 & \dd_1 \\
0 & 1
\end{pmatrix}\,.
$$
In this way one obtains the second local factorization relation
(\ref{trick'1/2}).

The approach proposed here reproduces the local formulae
(\ref{trick1/2}), (\ref{trick'1/2}) and in some sense explains the
origin of these peculiar relations on which the construction
in~\cite{Shortcut}
   relies: finally everything is based on the three-term
relations (\ref{rf1'}), (\ref{rf2'}). On the other hand our
construction in part I  is solely based on (\ref{rf1'}),
(\ref{rf2'}). This means, that both constructions are equivalent
generally, because they can be founded on the same basis.
However each of them has specific properties.

\subsection{Finite-dimensional operators $\mathbf{R},\,\mathbf{R}^{+}, \mathbf{R}^{-}$}

\label{RFinDim}

In the previous subsection we have demonstrated
the restriction of operators $\R,\,\R^{+},\,\R^{-}$ acting in the
space $\U_{-\ell}\otimes\U_{-s}$ to
invariant subspace $\V_{n}\otimes\U_{-s}$ in
the case $n=1$. For integer values $2 \ell = n$ the module
$\U_{-\ell}$ becomes reducible and the action of these three operators on
the first factor space becomes reducible.
This  can be checked by
analyzing the limits $\varepsilon \to 0, \ 2 \ell =
n - \varepsilon$ in the products $ \mathrm{R} = \mathrm{R}^{1}\,
\mathrm{R}^{2}$ (\ref{Rfact}) , $\mathrm{R}^{+} = \mathrm{r}^{+} \,\mathrm{R}^2$ (\ref{R+}) ,
$\mathrm{R}^{-} = \mathrm{R}^1 \,\mathrm{r}^{-}$ (\ref{R-}).

In the following it is convenient to use the projection operators
\begin{equation}
\label{proj} \Pi^n_i z_i^k = z_i^k\ ,\  k\le n\ \ \ ; \ \ \
\Pi^n_i z_i^k = 0\ ,\ k > n.
\end{equation}
Thus we restrict the operators as follows

\be \label{relation0} \mathbf{R}_{12}(u_1 , u_2|v_1,v_2) =
\lim_{\varepsilon\to 0} \,
\R_{12}(u_1+\textstyle{\frac{\varepsilon}{2}} ,
u_2-\textstyle{\frac{\varepsilon}{2}}|v_1,v_2)\,\Pi_1^n \ee

\begin{equation}\label{relation1}
\mathbf{R}^{+}_{12}(u_1,u_2 |v_2) =  \lim_{\varepsilon\to 0}
\,\R^{+}_{12}(u_1+\textstyle{\frac{\varepsilon}{2}},
u_2-\textstyle{\frac{\varepsilon}{2}}|v_2)\,\Pi_1^n
\end{equation}

\begin{equation}\label{relation2}
\mathbf{R}^{-}_{12}(u_1,u_2 |v_1) = \lim_{\varepsilon\to 0}
\,\R^{-}_{12}(u_1+\textstyle{\frac{\varepsilon}{2}} ,
u_2-\textstyle{\frac{\varepsilon}{2}}|v_1)\,\Pi_1^n
\end{equation}
where the parameters are \be \label{paramfindim} u_1 = u -
{\textstyle\frac{n}{2}} -1\ , \ u_2 = u +
{\textstyle\frac{n}{2}}\quad ;\quad v_1 = v - s -1\ , \ v_2 = v +
s\ \ee We shall write in boldface style operators in integer or
half-integer spin cases restricted to the finite-dimensional
irreducible subspace. Let us mention that
$\mathbf{R},\,\mathbf{R}^{+},\,\mathbf{R}^{-}$ do not map beyond
subspace $\V_n\otimes\U_{-s}$ as it should be.

In Appendix C we calculate such restrictions to the space
$\V_{n}\otimes\U_{-s}$ of the $\R$-operator in general case, i.e.
at $2\ell \to n ,\ n= 0 , 1 , 2 ,\cdots $. We also specify there the
explicit expression for $\mathbf{R}^{+}$, the  restriction of $\R^{+}$, and
$\mathbf{R}^{-}$, the restriction of $\R^{-}$. We present them
as certain finite-dimensional differential operators since such a form
is convenient. Then
choosing particular values of $n$ one can construct the corresponding matrices
as we have done in the previous section.

In the Section~\ref{R+R-} performing reductions
in $\mathrm{RLL}$-relations we have introduced the
operators $\mathrm{R}^{+}$ and $\mathrm{R}^{-}$
as certain limits of the general $\mathrm{R}$-operator (\ref{R+}), (\ref{R-}).
One can perform similar reduction of the operator $\mathbf{R}$
cancelling the leading dependence on the large $v_1$ or $v_2$
like in (\ref{R+}), (\ref{R-}).
In this way we obtain\footnote{
The relations (\ref{bfR+}) and (\ref{bfR-})  should be understood
up to normalization which we fix as in (\ref{bfR+Expl}) and (\ref{bfR-Expl})}
once more the operators
$\mathbf{R}^{+}$ and $\mathbf{R}^{-}$,
\begin{equation}\label{bfR+}
v_1^{z_1\partial_1}\,\mathbf{R}(u_1,u_2|v_1,v_2) \to
\mathbf{R}^{+}(u_1,u_2|v_{2}) \qquad  \text{at} \quad v_1 \to \infty,
\end{equation}

\begin{equation}\label{bfR-}
\mathbf{R}(u_1,u_2|v_1,v_2)\,
v_2^{-z_1\partial_1} \to \mathbf{R}^{-}(u_1,u_2|v_{1})
\qquad  \text{at} \quad v_2 \to \infty .
\end{equation}
This shows that we can obtain $\mathbf{R}^{+}$ and $\mathbf{R}^{-}$ in two ways:
at first perform the reduction of the general
$\R$-operator taking one of its parameters to infinity
and then restrict to the finite-dimensional subspace or
inversely at first restrict to the subspace and only then
perform the reduction taking the appropriate limit.
 (\ref{bfR+}), (\ref{bfR-}) can be obtained
using Stirling's formula (\ref{Stirling})
and performing calculation analogous to the one in Appendix A.

\subsection{The general transfer matrices and $\mathbf{Q}_{\pm}$-operators}

After the necessary preparation in the previous section we are going to
the construction of the general transfer matrix (\ref{T})
out of the new building blocks $\mathbf{R}_{k0}$
if $\ell$ is (half)-integer
\be \label{Tfin'}
\mathbf{T}_{s}(u) = \tr_0  \, q^{z_0 \dd_0} \,
\mathbf{R}_{10}\left(u|{\textstyle\frac{n}{2}}, s\right)
\mathbf{R}_{20}\left(u|{\textstyle\frac{n}{2}}, s\right) \cdots
\mathbf{R}_{N0}\left(u|{\textstyle\frac{n}{2}}, s\right) \, \ee
which is well defined on the finite-dimensional quantum space of
the model. In this case the trace over infinite-dimensional auxiliary
space $\C[z_0]$ diverges without $q$-regularization.


Now we consider the factorization of the general transfer
matrix (\ref{Tfin}). We start as before with the three term relation
(\ref{rf1'}) and restrict it on site $k$  to
$\V^n\otimes \C[z_0]\otimes \C[z_{0'}]$ at
$\ell=\frac{n}{2}$
$$
\R^1_{00^{\prime}}(v_1|w_1,w_2) \,
\mathbf{R}_{k0^{\prime}}(u_{1},u_2|w_1,w_2) \,
\mathbf{R}_{k0}(u_{1},u_2|v_1,v_{2}) =
$$
\be \label{rf1''}
=\mathbf{R}_{k0}(u_{1},u_2|v_1,w_{2}) \,
\mathbf{R}_{k0^{\prime}}(u_1,u_{2}|w_1,v_{2}) \,
\R^1_{00^{\prime}}(v_1|w_1,w_2)
\ee
Then we have to do the
 appropriate limiting procedure in the previous relation
taking at first $w_1 \to \infty$
and then $w_2 \to \infty$
This leads to
$$
\mathrm{P}_{00^{\prime}} \,
(-)^{z_{0^{\prime}}\partial_{0^{\prime}}}\,
e^{z_0\partial_{0^{\prime}}} \cdot
e^{z_{0^{\prime}} \dd_k } \,\Pi^n_k
\cdot \mathbf{R}_{k0}(u_{1},u_2|v_1,v_{2}) =
$$
\be \label{BFlocalT->-+} = \mathbf{R}^{-}_{k0}(u_{1},u_2|v_1)\cdot
\mathbf{R}_{k0^{\prime}}^{+}(u_1,u_{2}|v_{2})\cdot
\mathrm{P}_{00^{\prime}} \,
(-)^{z_{0^{\prime}}\partial_{0^{\prime}}} \,
e^{z_0\partial_{0^{\prime}}} \ee
We present the derivation of
(\ref{BFlocalT->-+}) in Appendix C. This local relation leads in the
standard way to the factorization relation for corresponding
regularized transfer matrices \be \label{T->Q+Q-Fin}
\frac{1}{1-q}\cdot\mathbf{T}_{s} (u) = \mathbf{Q}_{+} (u - s) \,
\mathbf{Q}_{-} (u + s + 1) = \mathbf{Q}_{-} (u + s + 1)\,
\mathbf{Q}_{+} (u - s) \, \ee
where $\mathbf{Q}_{-},\,\mathbf{Q}_{+}$ are transfer matrices
constructed from building blocks $\mathbf{R}^{-}_{k0},\,\mathbf{R}^{+}_{k0}$
\be \label{bfQ-}
 \mathbf{Q}_{-}(u-v_1) = \tr_{0}
\, q^{z_0 \dd_0} \, \mathbf{R}^{-}_{10}(u_1,u_2|v_1)\cdots
\mathbf{R}^{-}_{N0}(u_1,u_2|v_1)\,,
\ee
\be \label{bfQ+}
\mathbf{Q}_{+}(u-v_2) = \tr_{0} \, q^{z_0 \dd_0} \,
\mathbf{R}^{+}_{10}(u_1,u_2|v_2)\cdots
\mathbf{R}^{+}_{N0}(u_1,u_2|v_2)\,
\ee
The second variant of the factorization in the above relation can be obtained similarly from
appropriate local relation.

This construction is analogous to the
one in Section 4, where $\ell$ was a generic
complex number. The proof of the commutativity \be
\label{commutQ+Q-Fin} [ \, \mathbf{T}_{s}(u) , \mathbf{Q}_{\pm} (v) \,
] = 0 \ \ ;\ \ [ \, \mathbf{Q}_{\pm} (u) , \mathbf{Q}_{\pm} (v)\, ] =  0 \
\ ;\ \ [ \, \mathbf{Q}_{-} (u) , \mathbf{Q}_{+}(u) \, ]  = 0
\ee uses the
general Yang-Baxter equation and also goes parallel the corresponding
derivation given in Section 4.

Let us stress once more  that we have based each relation
for the general transfer matrix and Baxter operators
on a corresponding local relation for their building blocks.
We derive such local relations from  three basic
relations (\ref{rf'}), (\ref{rf1'}), (\ref{rf2'})
performing appropriate limits.
The general Yang-Baxter relation (\ref{rf'}) produces
local relations implying commutativity of the corresponding transfer matrices
and (\ref{rf1'}), (\ref{rf2'}) produce local relations implying factorization
relations for transfer matrices. Considering the construction with finite-dimensional
representations in the quantum space we at first restrict the basic
relations (\ref{rf'}), (\ref{rf1'}), (\ref{rf2'}) to the finite-dimensional subspace
in quantum space and only then perform appropriate limiting procedures which exclude
a part of the parameters. In this way, for example, we obtain
the sequence of relations (\ref{rf1'}) $\to$ (\ref{rf1''}) $\to$ (\ref{BFlocalT->-+}).

We have shown that the local operators $\R^{+}$ and $\R^{-}$ admit the
restriction to the finite-dimensional subspace in quantum space
(\ref{relation1}), (\ref{relation2}) if $\ell$ is (half)-integer.
Consequently the same is valid  for the Baxter operators
 constructed out of them.  Indeed, we have in the limit
$\varepsilon \to 0,\,2 \ell = n - \varepsilon$
\begin{equation}\label{bfQ+-}
\mathbf{Q}_{\pm}(u) =  \lim_{\varepsilon\to 0}
\,\mathrm{Q}_{\pm}(u)\bigl|_{\ell =
\frac{n-\varepsilon}{2}}\,\cdot\,\Pi^n , \qquad  \text{where} \qquad
\Pi^n \equiv \Pi^n_1 \Pi^n_2 \cdots \Pi^n_N .
\end{equation}

Before  discussing the Baxter equation
we derive some formulae connecting Baxter
operators $\mathbf{Q}_{\pm}$ introduced above with
Baxter operators $\mathbf{Q}_{1,2}$ constructed in part I.
This connection implies the form of the
Baxter equations for operators $\mathbf{Q}_{\pm}$.

\subsection{The operators $\mathbf{R}^1,\,\mathbf{R}^2,\,\mathbf{S}$}
\label{R1R2Sfindim}

Along with the operator $\mathbf{R}(u_1,u_2|v_1,v_2)$ we have also
introduced in part I the operators $\mathbf{R}^1(u_1|v_1,v_2)$
and $\mathbf{R}^2(u_1,u_2|v_2)$ where parameters $u_1,\,u_2,\,v_1,\,v_2$
respect the relation (\ref{paramfindim}), i.e. the quantum
spin $\ell$ is (half)-integer.
They are obtained from the former one by imposing additional relations on parameters.
Taking the limit $\delta\to0$ we obtain
\be \label{bfR1}
\mathbf{R}_{12}(u_1 , u_2|v_1,u_2-\delta) \to
\delta^{-1}\cdot\mathbf{R}^1_{12}(u_1 |v_1,u_2)\,,
\ee

\be \label{bfR2}
\mathbf{R}_{12}(u_1 , u_2|u_1+\delta,v_2) \to
\delta\cdot\mathbf{R}^2_{12}(u_1, u_2 |v_2)\,,
\ee

\begin{equation}\label{Rfindim}
\mathbf{R}_{12}(u_1 , u_2|u_1 + \delta ,u_2 - \delta ) \to \mathbf{S}_{12} \equiv
\mathrm{P}_{12}\cdot e^{-z_1 \dd_2}\,\Pi^n_2\, e^{z_1 \dd_2} \cdot
\Pi_1^n 
\end{equation}

The operators $\mathbf{R}^{1}$ and $\mathbf{R}^2$ are finite-dimensional analogues
of $\R^1$ and $\R^2$ (\ref{R1R2}).
They do not map beyond
$\V_{n}\otimes\U_{-s}$. Indeed they are special
limits of $\mathbf{R}$ which is well defined on the finite-dimensional
subspace.
Let us mention that the operator $\R^2$ does not admit a restriction
to the finite-dimensional subspace, i.e. it maps beyond $\V_{n}\otimes\U_{-s}$.
It is the main reason for constructing building the blocks for transfer matrices
from $\mathbf{R}$ instead of taking them to be $\R^{1}$ or $\R^2$.

The operator $\mathbf{S}_{12}$
is a nontrivial analogue of the transposition operator $\mathrm{P}_{12}$
for the finite-dimensional construction.
It also does not map beyond  $\V_{n}\otimes\U_{-s}$. Let us recall that
$\R_{12}(u_1,u_2|u_1,u_2) = \mathrm{P}_{12}$ in contrast to (\ref{Rfindim}).
This example clearly shows that the order
in which we impose the relations on parameters and take the spin in quantum space
$\ell$ to be (half)-integer is important.

In the Section~\ref{r+r-} we have considered the operators
$\mathrm{r}^{+}$ and $\mathrm{r}^{-}$
which are certain reductions of $\mathrm{R}^1$ and $\mathrm{R}^2$.
Now we are going to introduce their analogues
for the finite-dimensional case, i.e. we perform the
reduction of operators $\mathbf{R}^1$ and $\mathbf{R}^2$.
We adopt here the same rule (\ref{r-}), (\ref{r+}) to
perform the limiting procedure,
\be \label{bfr+}
v_1^{z_1\partial_1}\,\mathbf{R}^{1}(u_1|v_1,v_2) \to
\mathbf{r}^{+}(u_1|v_{2}) \qquad  \text{at} \quad v_1 \to \infty ,
\ee

\be\label{bfr-}
\mathbf{R}^{2}(u_1,u_2|v_2) \,
v^{-z_1 \dd_1}_2 \to \mathbf{r}^{-}(u_1|u_2)
\qquad  \text{at} \quad v_2 \to \infty .
\ee

Explicit expression for the operators $\mathbf{R}^{1},\,\mathbf{R}^2,\,\mathbf{r}^{-}$
and $\mathbf{r}^{+}$
can be found in Appendix C.

\subsection{Relations between $\mathbf{Q}_{\pm}$ and $\mathbf{Q}_{1,2}$}

In Section 4 we have pointed out the connection between
two constructions of Baxter operators for infinite-dimensional
representations. Now we are going to do the same for the
finite-dimensional case.
In part I (compare Appendix C there)
we have introduced the $q$-regularization into the traces for
Baxter operators $\mathbf{Q}_1$ and $\mathbf{Q}_2$.
$$ \mathbf{Q}_1(u-v_1) = \tr_{0}
\, q^{z_0 \dd_0} \, \mathbf{R}^1_{10}(u_1|v_1,u_{2})\cdots
\mathbf{R}^1_{N0}(u_1|v_1,u_{2})\,,
$$
$$
\mathbf{Q}_2(u-v_2) = \tr_{0} \, q^{z_0 \dd_0} \,
\mathbf{R}^2_{10}(u_1,u_2|v_{2})\cdots
\mathbf{R}^2_{N0}(u_1,u_2|v_{2})\,
$$
\be \label{Sbold} \mathbf{S} = \tr_{0} \, q^{z_0\dd_0} \,
\mathbf{S}_{10} \, \mathbf{S}_{20} \cdots \mathbf{S}_{N0} \,,\ee
They obey the factorization relation
\be
\label{T->Q2Q1Fin} \mathbf{S}\,\mathbf{T}_{s} (u) = \mathbf{Q}_{2}
(u - s) \, \mathbf{Q}_{1} (u + s + 1) = \mathbf{Q}_{1} (u + s +
1)\, \mathbf{Q}_{2} (u - s) \, \ee
very similar to (\ref{T->Q+Q-Fin}) and satisfy the Baxter equations
$$
\mathrm{t}(u|q)\mathbf{Q}_1(u) =  \mathbf{Q}_1(u+1)+ q\cdot(u_1
u_2)^N\cdot\mathbf{Q}_1(u-1)\, ,
$$
\begin{equation}\label{finBaxter2}
\mathrm{t}(u|q) \, \mathbf{Q}_2(u) = q \cdot \mathbf{Q}_2(u+1)+
(u_1 u_2)^N \cdot \mathbf{Q}_2(u-1)\, .
\end{equation}

We also need the auxiliary transfer matrices involving
$\mathbf{r}^{\pm}$ (\ref{bfr+}), (\ref{bfr-}) of the form
\be \label{bfq+}
\mathbf{q}_{+} = \tr_{0} \left[q^{z_{0} \dd_{0} } \,
\mathbf{r}^{+}_{10}(u_1|u_2)\cdots
\mathbf{r}^{+}_{N0}(u_1|u_2)\right] \,, \ee \be
\label{bfq-} \mathbf{q}_{-} = \tr_{0} \left[q^{z_{0} \dd_{0} } \,
\mathbf{r}^{-}_{10}(u_1|u_2)\cdots
\mathbf{r}^{-}_{N0}(u_1|u_2)\right]\,. \ee
For brevity the dependence on $\ell$ and on the parameter of regularization $q$ is
suppressed in the notation.
Choosing the parameters in (\ref{T->Q+Q-Fin})
in a special way we obtain
\be \label{bfq-q+} \frac{1}{1-q}\cdot\mathbf{S} =
\mathbf{q}_{-} \cdot \mathbf{q}_{+} = \mathbf{q}_{+} \cdot
\mathbf{q}_{-}
\ee

In order to connect  the Baxter operators $\mathbf{Q}_{1}$ and $\mathbf{Q}_{-}$
we need the local factorization relation
$$
\R^1_{00^{\prime}}(v_1|w_1,u_2) \cdot
\mathbf{R}^{1}_{k0^{\prime}}(u_{1}|w_1,u_2) \cdot
\mathbf{r}_{k0}^{-}(u_{1}|u_2) =
$$
$$
=\mathbf{S}_{k0} \cdot \mathbf{R}^{-}_{k0^{\prime}}(u_1,u_{2}|w_1) \cdot
\R^1_{00^{\prime}}(u_1|w_1,u_2)\,
$$
It produces the corresponding relation for the transfer matrices.
Thus we obtain
\be \label{bfQ-->Q1q-} \mathbf{Q}_{-}(u) =
\mathbf{S}^{-1} \cdot \mathbf{Q}_{1}(u) \cdot \mathbf{q}_{-}
\ee 
and similarly
\be \label{bfQ+->q+Q2}
\mathbf{Q}_{+}(u) = \mathbf{S}^{-1} \cdot \mathbf{q}_{+} \cdot
\mathbf{Q}_{2}(u)
\ee
It is also
possible to deduce commutativity
of the operators $\mathbf{q}_{\pm},\,\mathbf{Q}_{\pm},\,\mathbf{Q}_{1},\,\mathbf{Q}_{2}$
and $\mathbf{S}$ among themselves from local intertwining relations for their building blocks.
As we have mentioned above such local relations underlying
factorization and commutativity can be obtained from
(\ref{rf'}), (\ref{rf1'}), (\ref{rf2'}) restricted to
finite-dimensional subspace. Further comments can be found in Appendix C.

The Baxter equations (\ref{finBaxter2}) for $\mathbf{Q}_{1,2}$ imply
the ones for $\mathbf{Q}_{\pm}$
\be \label{finBaxter-}
\mathrm{t}(u|q)\,\mathbf{Q}_{-}(u) =\mathbf{Q}_{-}(u+1) + q\cdot(u_1
u_2)^N \cdot\mathbf{Q}_{-}(u-1) \,,
\ee
\be \label{finBaxter+} \mathrm{t}(u|q) \,
\mathbf{Q}_{+}(u) = q\cdot \mathbf{Q}_{+}(u+1) + (u_1 u_2)^N
\cdot\mathbf{Q}_{+}(u-1) \,.
\ee
However they can also be obtained in the other way
(see comments in Section~\ref{gen_tm}),
i.e. there are local relations which produce the Baxter equations.


\subsection{Explicit action on polynomials}

\label{explicitfindim}

Now we proceed to calculate explicitly the action of the
 Baxter operators $\mathbf{Q}_{+},\,\mathbf{Q}_{+}$
 on polynomials.
In  Section~\ref{explicit} we have already computed the
action of the Baxter operators $\mathrm{Q}_{+},\,\mathrm{Q}_{-}$
constructed for infinite-dimensional representation to polynomials
and have obtained explicit formulae. We are going to use
these results and to take into account (\ref{bfQ+-}), i.e.
we only have to consider the
polynomials from appropriate subspace,
apply the formulae from Section~\ref{explicit} and
do the limit $\ell\to\frac{n}{2}$.

Thus we obtain immediately the explicit formula for the renormalized $\mathbf{Q}_{+}$
using (\ref{Q(+)pol})
\begin{equation}\label{bfQ(+)pol}
\mathbf{Q}^{+}(u) \, : \, (1-x_{1}z_{1})^{n} \cdots (1-x_{N}z_{N})^{n}\, \mapsto \,
\end{equation}
$$
\mapsto \, \overline{\Pi}^n \cdot
\prod^{N}_{k = 1} \left( 1 - \frac{x_k}{1-q}\cdot( z_{1,k-1} + q
z_{k,N} ) \right)^{\frac{n}{2}-u} \left( 1 - \frac{x_k}{1-q}\cdot(
z_{1,k} + q z_{k+1,N} ) \right)^{\frac{n}{2}+u}\,,
$$
where
$$
z_{1,k} \equiv z_1 + z_2 + \cdots + z_k \qquad ; \qquad z_{k,N} \equiv z_k
+ z_{k+1} + \cdots + z_N \qquad ;
\overline{\Pi}^n \equiv \overline{\Pi}^n_1 \overline{\Pi}^n_2
\cdots \overline{\Pi}^n_N
$$
and $\overline{\Pi}^n_k \equiv \lim_{\varepsilon \to 0}
\frac{\Gamma( z_k \dd_k - n+\varepsilon)}{\Gamma(-n+\varepsilon)}
= \frac{n!\,(-1)^{z_k \dd_k}}{\Gamma(1+n-z_k \dd_k)} \cdot \Pi^n_k$.

Then we turn to the second Baxter operator. We use
(\ref{Q-->Q1q-}) and rearrange there $\varepsilon$
in such a way that the limits $\varepsilon\to0,\, 2\ell = n - \varepsilon$ exist
for each operator factor,
\be \mathrm{Q}_{-}(u) =
\varepsilon^N \mathrm{Q}_{1}(u|q) \cdot
\frac{1}{\varepsilon^N} \, q^{-z_1 \dd_1} \mathrm{P}^{-1} \, \mathrm{q}_{-}
\ee
From (\ref{q-factor}), (\ref{q-trace}) we obtain
\be \label{q-pol}
\frac{1}{\varepsilon^N}\cdot q^{-z_1 \dd_1} \mathrm{P}^{-1} \cdot \mathrm{q}_{-} \, \Psi(z_1, \cdots , z_N ) \to
\Psi( z_2 - z_1, z_3 - z_2 , \cdots ,
q^{-1} z_{1}- z_N)
\ee
where $\Psi(\vec{z}\,)$ is a polynomial from the
finite-dimensional subspace: $\Pi^n \, \Psi(\vec{z}\,) = \Psi(\vec{z}\,)$.

In the limit $\varepsilon\to0$ explicit the expression for operator $\mathrm{Q}_1(u|q)$ (\ref{bfQ1q})
becomes much  simpler. In  part I we have obtained
\begin{equation}\label{Q1findim}
\varepsilon^N\cdot\mathrm{Q}_{1}(u|q)\, \Phi(\vec{z} ) \to \frac{1}
{\Gamma^N(1+\frac{n}{2}-u)\,n!^N}\cdot\dd^n_{\lambda_1}\cdots
\dd^n_{\lambda_N} \left. \frac{\left(\bar{\lambda}_1 \cdots
\bar{\lambda}_N\right)^{\frac{n}{2}-u} }{ 1 - q \bar{\lambda}_1
\cdots \bar{\lambda}_N } \cdot \Phi(\Lambda_{q}^{\prime -1}
\vec{z} \, ) \right|_{\lambda = 0} \,
\end{equation}
where $\Phi(\vec{z}\,)$ is an arbitrary polynomial and $\bar{\lambda}\equiv 1-\lambda$.

Combining (\ref{q-pol}) with (\ref{Q1findim}) and changing normalization
we obtain finally the explicit formula
\be
\mathbf{Q}^{-}(u)\, \Psi(\vec{z} ) = \dd^n_{\lambda_1}\cdots
\dd^n_{\lambda_N} \left. \frac{\left(\bar{\lambda}_1 \cdots
\bar{\lambda}_N\right)^{\frac{n}{2}-u} }{ 1 - q \bar{\lambda}_1
\cdots \bar{\lambda}_N } \cdot \Psi(\Lambda_{q}^{\prime -1} \mathrm{M}
\vec{z} \, ) \right|_{\lambda = 0} \,.
\ee
where
$$
\mathrm{M} = \begin{pmatrix}
-1 & 1 & 0 & \hdotsfor{2} & 0 \\
0 & -1 & 1 & \hdotsfor{2} & 0 \\
\hdotsfor{6}  \\
0 & 0 & \hdotsfor{2} & -1 & 1 \\
\frac{1}{q} & 0 & 0 & \cdots & 0 &
-1
\end{pmatrix} \ ; \
\Lambda_q^{\prime} = \begin{pmatrix}
1-\frac{1}{\lambda_1} & \frac{1}{\lambda_1} & 0 & 0 & \hdotsfor{2} & 0 \\
0 & 1-\frac{1}{\lambda_2} & \frac{1}{\lambda_2} & 0 & \hdotsfor{2} & 0 \\
0 & 0 & 1-\frac{1}{\lambda_3} & \frac{1}{\lambda_3} & \hdotsfor{2} & 0 \\
\hdotsfor{7}  \\
0 & 0 & \hdotsfor{3} & 1-\frac{1}{\lambda_{N-1}} & \frac{1}{\lambda_{N-1}} \\
\frac{1}{q \lambda_{N}} & 0 & 0 & 0 & \cdots & 0 &
1-\frac{1}{\lambda_{N}}
\end{pmatrix}
$$

\section{Discussion}

In two papers we have studied in  great detail two variants of the
construction of $\mathrm{Q}$-operators in the framework of
QISM~\cite{FST,TTF,KuSk1,Fad}. We use the main ingredients of QISM,
the  $\mathrm{L}$-matrices, the general $\mathrm{R}$-matrices and the local
commutation relations between them. As stated in
Introduction the general scheme consists of two main steps:

First of all one needs the general or universal
$\mathrm{R}$-matrix -- the general solution of Yang-Baxter
equation. The general $\mathrm{R}$-matrix can be factorized in a
product of simpler operators which  serve as building blocks
for $\mathrm{Q}$-operators.

The second step is the proof of the factorization of the general
transfer matrix, constructed from the general $\mathrm{R}$-matrix
into the product of $\mathrm{Q}$-operators. This factorization
allows to derive the Baxter equation for $\mathrm{Q}$-operators and
various fusion relations.

We have  unified two approaches by deducing all relations for
transfer matrices and Baxter operators from local relations for
their building blocks in the spirit of QISM. The local relations
in turn are derived from three basic local relations of
Yang-Baxter type. We consider this simplification of argumentation
as one of the main achievements. It allows to reproduce the known
results in much more transparent and systematical way.

Further, we have included in one scheme the treatment of
infinite-dimensional representations in the quantum space as well as all
finite-dimensional representations.

We believe that the main goal is not achieved by having done
the  pure algebraic construction of $\mathrm{Q}$-operators as
transfer matrices but includes also the explicit description of these operators.
Therefore  we have derived explicit formulae for the action of
$\mathrm{Q}$-operators on polynomials representing the quantum states of the
chain.

Let us discuss now the particular results obtained in the present paper.
Comparing the ordinary transfer matrix $\mathrm{t}(u)$, its
generalizations $\mathrm{t}_n(u), \mathrm{T}_s(u)$ and the Baxter
operators  a simple systematics in their
construction is evident. All they are constructed as traces of products of
operators with one factor for each chain site including the
$q$-regularization of the traces. The factor operators act on the
tensor product of the quantum and the auxiliary spaces. In the
construction with infinite-dimensional representations in the
quantum space at generic spin $\ell$ in the most general case the
factor at site $k$ is the general Yang-Baxter operator $\R_{k0}$.
In the other cases the factor operators are certain reductions
obtained therefrom by imposing conditions on the representation
parameters $v_1, \,v_2$ referring to the auxiliary space and/or
extracting the asymptotics for  representation
parameters $v_1, \,v_2$ to infinity:
\begin{center}
\begin{tabular}{ccc}
{\it chain operator} \ \ & {\it site operator}\ \ \  & {\it restriction} \cr
 $\mathrm{T}_s$ & $\R_{k0}$                   & --- \cr
$\mathrm{t} $    & $\mathrm{L}_k \sim \mathbf{R}_{k0}(\ell,\frac{1}{2})$&
$v_2-v_1 = 2$ and  $\Pi_0^1$ \cr
$\mathrm{Q}_+ $ & $\R^+_{k0}$ & $v_1 \to \infty$ \cr
$\mathrm{Q}_- $ & $\R^-_{k0}$ & $v_2 \to \infty$
\end{tabular}
\end{center}

\noindent There are also the cases of twofold restrictions, where
conditions are imposed on both $v_1$ and $v_2$ resulting in
simpler operators:

\begin{center}
\begin{tabular}{ccc}
{\it chain operator} \ \ & {\it site operator} \ \  & {\it restrictions} \cr
$\mathrm{q}^+$ & $\mathbb{r}^{+}_{k0}$  & $v_2 = u_2 \ , \ v_1 \to
\infty$ \cr
$\mathrm{q}^-$ & $\mathbb{r}^{-}_{k0}$ & $ v_1 = u_1\ , \ v_2 \to
\infty$ \cr
$\mathrm{P}\,q^{z_1 \dd_1}$ & $\mathrm{P}_{k0}$  & $ v_1 = u_1\ , \ v_2 = u_2 $ \cr
\end{tabular}
\end{center}

\noindent The general transfer matrix $\mathrm{T}_s$ factorizes
into the products of $\mathrm{Q}_+$ and $\mathrm{Q}_-$. The proofs
of factorization and commutativity for different transfer matrices
rely on local three-term relations –- Yang-Baxter relations.

The construction in~\cite{Shortcut} is based on local relations
including $\mathrm{L}$-operators acting in the tensor product of the
two-dimensional quantum space and the  infinite-dimensional auxiliary
space. We work with the general $\R$-operator and its reductions
acting in the tensor product of two infinite-dimensional spaces
$\U_{-\ell}\otimes\U_{-s}$ and formulate for them appropriate
local relations. That means we perform the lifting to the level of
$\R$-operators.  In particular
we  have shown that their reduction to
two-dimensional subspace in quantum space at $\ell = \frac{1}{2}$,
i.e. descending to the level of $\mathrm{L}$-operators, produces the
local relation of~\cite{Shortcut}.
This step of lifting provides the advantage of the unified
treatment of the cases of generic and all finite-dimensional
representations.

The systematics in the relation between global chain operators and local
building blocks applies in analogy also to the case of integer or
half-integer spin $\ell = \frac{n}{2}$ with finite-dimensional
representation spaces at the sites. In the  case of the general
transfer matrix $\mathbf{T}_s$
 the site operators are now $\mathbf{R}_{k0}(u|\frac{n}{2}, s)$,
 the Yang-Baxter operators
restricted to the irreducible subspace by means of the projector
$\Pi^n_k$ at $u_2-u_1 = n+1$. Additional restrictions on the parameters
$v_1, v_2$ lead to the other reductions:
\begin{center}
\begin{tabular}{ccc}
{\it chain operator} \ \ & {\it site operator}\ \ \  & {\it additional restriction} \cr
$\mathbf{T}_s$ & $\mathbf{R}_{k0}$                   & --- \cr
$\mathbf{Q}_{+}$ & $\mathbf{R}^{+}_{k0}$ & $v_1 \to \infty$ \cr
$\mathbf{Q}_{-} $ & $\mathbf{R}^{-}_{k0}$ & $v_2 \to \infty$
\end{tabular}
\end{center}

The presented proofs of factorization and commutativity of transfer matrices
for finite-dimensional representations are completely parallel to the
corresponding proofs of the infinite-dimensional case.
Our analysis clarifies the relation between the cases of generic spins
and half-integer or integer spins: by taking the limit $\ell\to\frac{n}{2}$
in the Baxter operators $\mathrm{Q}_{\pm}$ we obtain immediately the appropriate
set of Baxter operators $\mathbf{Q}_{\pm}$ which do not map beyond
the  finite-dimensional subspace, i.e. Baxter operators $\mathrm{Q}_{\pm}$
admit the restriction to the finite-dimensional subspace as well as the general
transfer matrix $\mathrm{T}_s$ does.
The reason lies in the property shared by their building blocks
$\R^{+},\,\R^{-},\,\R$
which all admit the restriction to finite-dimensional subspace
in quantum space of the site.

Moreover, besides presenting general formulae for
Baxter operators and proving their algebraic properties
we  present explicit compact expressions for their action on polynomials.

Another important achievement consists in establishing the relations
between the present construction following the scheme of \cite{Shortcut}
to the construction considered in part I. There we have
constructed Baxter operators
$\mathrm{Q}_{1,2}$ for infinite-dimensional representations and
Baxter operators $\mathbf{Q}_{1,2}$ for finite-dimensional
representations and have proven the corresponding factorization
formulae for the general transfer matrix. The unification of both
constructions and the detailed comparison was the main aim of the
present paper.

Let us start this discussion with the case of infinite-dimensional
representations. We have shown
that $\mathrm{Q}_{-}$ and $\mathrm{Q}_{+}$ differ from
$\mathrm{Q}_{1}$ and $\mathrm{Q}_{2}$ essentially by some dressing
factors $\mathrm{q}_{-}$ or $\mathrm{q}_{+}$, which are roughly
speaking inverse to each other. The operators $\mathrm{q}_{\pm}$ do
not depend on spectral parameter and play a passive role.
Indeed we have derived Baxter equations for $\mathrm{Q}_{\pm}$
straightforwardly from the ones for $\mathrm{Q}_{1,2}$. That is
both sets of Baxter operators $\mathrm{Q}_{1,2}$ and
$\mathrm{Q}_{\pm}$ contain identical information about the quantum
system but the operators $\mathrm{Q}_{1,2}$ are simpler. Moreover, the
essential shortcoming of the construction of
$\mathrm{Q}_{\pm}$-operators is the unavoidable $q$-regularization
violating $s\ell_2$ symmetry. It appears in the factorization
relations $\mathrm{T} \sim \mathrm{Q}_{+} \cdot \mathrm{Q}_{-}$
and in the expression for $\mathrm{Q}_{+}$
even in the infinite-dimensional case, whereas the trace in
$\mathrm{Q}_{-}$ converges for generic $\ell$ without
$q$-regularization.
 The construction of Baxter operators $\mathrm{Q}_1$ and
$\mathrm{Q}_2$ does not need the $q$-regularization in the
infinite-dimensional case. It seems that the
construction of the operators $\mathrm{Q}_1$ and $\mathrm{Q}_2$ is
more adequate  in the infinite-dimensional case.

In the case of finite-dimensional representation the situation is
opposite.
First of all in this case the dressing factor $\mathrm{q}_{+}$
relating   $\mathrm{Q}_{+}$ to $\mathrm{Q}_{2}$ reveals its
significance: it ensures the invariance of the finite-dimensional
representation subspace under the action of $\mathrm{Q}_{+}$,
whereas $\mathrm{Q}_{2}$ does not have this property.
For this reason the connection between the Baxter operators
$\mathrm{Q}_{\pm}$ for generic spin
and $\mathbf{Q}_{\pm}$ for the ones for half-integer spin is simple:
we can just put $\ell = \frac{n}{2}$ and the restriction to the
irreducible subspace is straightforward. Comparing to part I we see
that there the situation is different as far as the relation between generic
and compact spin cases is concerned. We have emphasized there that the limit
$2\ell \to n$ has to be performed with care in particular in the cases
where  $\mathrm{Q}_{2}$ or the parameter restriction $v_1 = u_1$ is involved.
Because  in the limit to integer values of $2\ell$ complications are absent
in the case of $\mathrm{Q}_{\pm}$ this construction appears more
adequate in the finite-dimensional representation case.
To summarize, both pairs of $\mathrm{Q} $ operators are complementary
in some sense and each of them has its favourable case of representations.

In both papers
we have restricted ourselves to the simplest example of
the spin chain with the symmetry algebra of rank one. There are a
plenty of results in more general situations.
The approach of~\cite{BLZ} and~\cite{Shortcut} is generalized
in~\cite{BHK,NewChapter} to the cases of algebras of higher rank and
in~\cite{BaTs,Oscillator} to the case of superalgebras.
The another approach is also applicable in these situations:
see~\cite{DeMa1,DeMa2,DeMa3} for $\mathrm{SL}(n,\mathbb{C})$
and~\cite{BDKM} for superalgebras.

\section*{Acknowledgement}

We thank V.Tarasov, A.Manashov and G.Korchemsky for discussions
and critical remarks.

This work has been supported by Deutsche Forschungsgemeinschaft
(KI 623/8-1). One of us (D.C.) is grateful to Leipzig University and DAAD
for support. The work of D.C. is supported by the Chebyshev
Laboratory (Department of Mathematics and Mechanics,
St.-Petersburg State University) under RF government grant
11.G34.31.0026. The work of S.D is supported by the RFFI grants
11-01-00570-a, 11-01-12037, 09-01-12150.
The work of D.K. is supported by Armenian grant 11-1c028.


\appendix
\renewcommand{\theequation}{\Alph{section}.\arabic{equation}}
\setcounter{table}{0}
\renewcommand{\thetable}{\Alph{table}}

\section*{Appendices}

\section{Reductions of $\mathrm{RLL}$-relations}
\setcounter{equation}{0}

Here we consider the calculation of reductions in
$\mathrm{RLL}$-relations and derive explicit expressions for operators
$\mathrm{r}^{+}$ and $\mathrm{R}^{+}$.

\subsection{Reduction $\mathrm{R}^1 \rightarrow \mathrm{r}^{+}$}

We start from the defining equation for operator
$\mathrm{R}^{1}$ (\ref{R1}) and take the limit $v_1 \to \infty$ using (\ref{asymptL})
$$
\mathrm{R}^{1}(u_1|v_1,v_2) \, \mathrm{L}_{1}(u_1,u_2)\,
\mathrm{L}^{-}_{2}(v_2) \left(\begin{array}{cc}
v_1 & 0 \\
0& 1\end{array}\right)=
\mathrm{L}^{-}_{1}(u_2)\left(\begin{array}{cc}
v_1 & 0 \\
0& 1\end{array}\right)\mathrm{L}_{2}(u_1,v_2)\,
\mathrm{R}^{1}(u_1|v_1,v_2)
$$
or in equivalent form
$$
\mathrm{R}^{1}(u_1|v_1,v_2) \, \mathrm{L}_{1}(u_1,u_2)\,
\mathrm{L}^{-}_{2}(v_2)  = \mathrm{L}^{-}_{1}(u_2)\,
\left(\begin{array}{cc}
v_1 & 0 \\
0&
1\end{array}\right)\mathrm{L}_{2}(u_1,v_2)\left(\begin{array}{cc}
v_1 & 0 \\
0& 1\end{array}\right)^{-1} \mathrm{R}^{1}(u_1|v_1,v_2) \ .
$$
The $s\ell_2$-invariance of the $\mathrm{L}$-operator allows to
transform the matrix similarity transformation to the operator similarity
transformation,
\begin{equation}\label{similar}
\left(\begin{array}{cc}
\lambda & 0 \\
0& 1\end{array}\right)\mathrm{L}(u)\left(\begin{array}{cc}
\lambda & 0 \\
0& 1\end{array}\right)^{-1} =
\lambda^{-z\partial}\,\mathrm{L}(u)\,\lambda^{z\partial} .
\end{equation}
Thus we obtain relation of the form (\ref{r+LL-})
\be \label{v1R1}
v_1^{z_2\partial_2}\,\mathrm{R}^{1}(u_1|v_1,v_2)\cdot
\mathrm{L}_{1}(u_1,u_2)\, \mathrm{L}^{-}_{2}(v_2)  =
\mathrm{L}^{-}_{1}(u_2)\, \mathrm{L}_{2}(u_1,v_2)\cdot
v_1^{z_2\partial_2}\, \mathrm{R}^{1}(u_1|v_1,v_2) \ . \ee
Now we
are going to calculate the leading term of $v_1^{z_2\partial_2}
\mathrm{R}^{1}(u_1|v_1,v_2)$ in asymptotics $v_1 \to \infty$
$$
v_1^{z_2\partial_2} \,\mathrm{R}^{1}(u_1|v_1,v_2) =
v_1^{z_2\partial_2}\cdot\frac{\Gamma(z_{21}\partial_2+u_1-v_2+1)}
{\Gamma(z_{21}\partial_2+v_1-v_2+1)} = v_1^{z_2\partial_2}\cdot
e^{-z_1\partial_2} \frac{\Gamma(z_{2}\partial_2+u_1-v_2+1)}
{\Gamma(z_{2}\partial_2+v_1-v_2+1)}e^{z_1\partial_2} =
$$
$$
= e^{-\frac{z_1\partial_2}{v_1}}
\frac{v_1^{z_2\partial_2}\cdot\Gamma(z_{2}\partial_2+u_1-v_2+1)}
{\Gamma(z_{2}\partial_2+v_1-v_2+1)}e^{z_1\partial_2}
\longrightarrow (2 \pi)^{\frac{1}{2}} \,
e^{v_1}\,v_1^{v_2-v_1-\frac{1}{2}}
\cdot\Gamma(z_{2}\partial_2+u_1-v_2+1)\,e^{z_1\partial_2}
$$
Above  we use Stirling's formula for the asymptotics of the
$\Gamma$-function
\be \label{Stirling} \Gamma(\lambda+a)
\overset{\lambda\to\infty}{\longrightarrow} (2 \pi )^{\frac{1}{2}}
e^{-\lambda}\,\lambda^{\lambda-\frac{1}{2}}\cdot\lambda^{a} . \ee
Note that the  factor $(2 \pi
)^{-\frac{1}{2}}\,e^{v_1}\,v_1^{v_2-v_1-\frac{1}{2}}$ can be
removed from both sides of equation~(\ref{v1R1})\footnote{In the
following we will omit such c-number factor and the notation
$v_1^{z_2\partial_2}\,\mathrm{R}^{1}(u_1|v_1,v_2) \to
\mathrm{r}^{+}(u_1|v_{2})$ implies it.} and
we obtain the operator $\mathrm{r}^{+}$ (\ref{r+}).
Thus we have specified the reduction procedure for the operator
$\mathrm{R}^1$ starting from our prescription for the reduction of the
$\mathrm{L}$-operator and using the $\mathrm{RLL}$ relation entangling
them.
That is, the  reduction procedure for
$\mathrm{L}$-operators determines the procedure for
$\mathrm{R}$-operators.

Similar calculations for the operator $\mathrm{R}^2$ leads to
(\ref{r-LL+}) and (\ref{r-}).

\subsubsection{Reduction $\mathrm{R} \rightarrow \mathrm{R}^{+}$}

Since the general $\mathrm{R}$-operator is constructed from
the operators $\mathrm{R}^1$ and $\mathrm{R}^2$ its reduction follows
from the building block reduction considered above.
We start from the defining equation for $\mathrm{R}$-operator
(\ref{RLL})
and take the limit $v_1 \to \infty$. Just in the same way as
before applying (\ref{similar}) we deduce \be \label{v1}
v_1^{z_2\partial_2}\,\mathrm{R}(u_1,u_2|v_1,v_2)\cdot
\mathrm{L}_{1}(u_1,u_2)\, \mathrm{L}^{-}_{2}(v_2)  =
\mathrm{L}^{-}_{1}(v_2)\, \mathrm{L}_{2}(u_1,u_2)\cdot
v_1^{z_2\partial_2}\, \mathrm{R}(u_1,u_2|v_1,v_2) \, . \ee
Calculating the leading term of
$$
v_1^{z_2\partial_2}\,\mathrm{R}(u_1,u_2|v_1,v_2) = v_1^{z_2\partial_2}
\,\mathrm{R}^{1}(u_1|v_1,u_2)\,\mathrm{R}^{2}(u_1,u_2|v_{2})
$$
in
asymptotic $v_1 \to \infty$
we see that the operator $\mathrm{R}^{2}(u_1,u_2|v_{2})$ does not
depend on $v_1$ and therefore it remains unchanged in the limit.
So the calculation of the previous section leads
to the operator $\mathrm{R}^+$ (\ref{R+})\footnote{ The notation
$\mathrm{R}^{+}$ and similarly for the reductions
of the $\mathrm{R}$-operators are chosen  in such a way that after
restriction to finite dimensional subspace we obtain
correspondingly $\mathrm{L}^{+}$.} which interchanges
$\mathrm{L}^{-}$ and $\mathrm{L}$ (\ref{R+LL-}).
Repeating the same procedure at $v_2 \to \infty$ we obtain
(\ref{R-LL+}) and (\ref{R-}).

\subsubsection{Double reductions}

Above we have implemented one reduction in the $\mathrm{RLL}$ relations
excluding  parameters by taking appropriate limits.
Now we perform one more reduction in the  $\mathrm{RLL}$ relations.

We take the limit $v_2 \to \infty$ in (\ref{R+LL-}) and obtain
$$ 
v_2^{-z_2 \dd_2} \, \mathrm{R}^{+}(u_1,u_2|v_{2}) \cdot
\mathrm{L}_{1}(u_1,u_2)\,
\left(\begin{array}{cc} 1 & 0 \\
      z_2 & 1 \end{array}\right) =
\left(\begin{array}{cc} 1 & 0 \\
      z_1 & 1 \end{array}\right) \,
\mathrm{L}_{2}(u_1,u_2)\cdot
v_2^{-z_2 \dd_2} \, \mathrm{R}^{+}(u_1,u_2|v_{2})
$$
Let us calculate the limit explicitly. At first we
rewrite $\mathrm{R}^{+}(u_1,u_2|v_{2})$ taking into
account (\ref{R+}), (\ref{r+}), (\ref{R1R2})
and the useful representation for the permutation,
$\mathrm{P}_{12}= e^{-z_2 \dd_1} \, e^{z_1 \dd_2} \,
e^{-z_2 \dd_1} \, (-)^{z_1 \dd_1}$,
$$
\mathrm{R}^{+}(u_1,u_2|v_{2}) = \mathrm{P}_{12} \,
\Gamma(z_{1}\partial_1+u_1-u_2+1) \, e^{z_1\partial_2} \,
(-)^{z_1 \dd_1}
\frac{\Gamma(z_{1}\dd_1+u_1-v_2+1)}{\Gamma(z_{1}\dd_1+u_1-u_2+1)}
\, e^{z_2\partial_1} .
$$
Thus using (\ref{Stirling}) one readily obtains
\be \label{v2R+}
v_2^{-z_2 \dd_2} \, \mathrm{R}^{+}(u_1,u_2|v_{2})
\to \mathrm{P}_{12} \, e^{z_2 \dd_1}
\ee

Then we calculate the limit $v_2 \to \infty$ in (\ref{r+LL-})
using (\ref{r+})
$$ 
v_2^{-z_2 \dd_2} \, \mathrm{r}^{+}(u_1|v_{2})
\cdot \mathrm{L}_{1}(u_1,u_2) \cdot
\left(\begin{array}{cc} 1 & 0 \\
      z_2 & 1 \end{array}\right) =
\mathrm{L}^{-}_{1}(u_2)\,
\mathrm{L}^{+}_{2}(u_1)\cdot
v_2^{-z_2 \dd_2} \, \mathrm{r}^{+}(u_1|v_{2})
$$
\be \label{2r+}
v_2^{-z_2 \dd_2} \, \mathrm{r}^{+}(u_1|v_{2}) \rightarrow
(-)^{z_2 \dd_2} e^{z_1 \dd_2}
\ee

It is also possible to calculate the reduction of (\ref{R+LL-}) at $u_1\to\infty$
\begin{equation}\label{R++L-L-}
\mathrm{R}^{++}(u_2|v_{2})\cdot \mathrm{L}^{-}_{1}(u_2)\,
\mathrm{L}^{-}_{2}(v_2)  = \mathrm{L}^{-}_{1}(v_2)\,
\mathrm{L}^{-}_{2}(u_2)\cdot \mathrm{R}^{++}(u_2|v_{2})
\end{equation}
\begin{equation}\label{R++}
\mathrm{R}^{+}(u_1,u_2|v_{2})\,u_1^{-z_2\partial_2} \to
\mathrm{R}^{++}(u_2|v_{2}) =
\left(1-z_2\partial_{1}\right)^{u_2-v_2}
\end{equation}
and the reduction of (\ref{R-LL+}) at $u_2\to\infty$
\begin{equation}\label{R--L+L+}
\mathrm{R}^{--}(u_1|v_{1})\cdot
\mathrm{L}^{+}_{1}(u_1)\, \mathrm{L}^{+}_{2}(v_1)  =
\mathrm{L}^{+}_{1}(v_1)\, \mathrm{L}^{+}_{2}(u_1)\cdot
\mathrm{R}^{--}(u_1|v_{1})
\end{equation}
\begin{equation}\label{R--}
u_2^{z_2\partial_2}\,\mathrm{R}^{-}(u_1,u_2|v_{1}) \to
\mathrm{R}^{--}(u_1|v_{1}) =
\left(1+z_1\partial_{2}\right)^{u_1-v_1}
\end{equation}
using similar methods as before.

\section{Local relations}

\setcounter{equation}{0}

Here we shall derive the local factorization and commutativity
relations which underlie the factorization of the general transfer
matrix and its commutativity.

\subsection{Local factorization}

We start with the derivation of (\ref{localT->-+}) which is the local
variant of (\ref{T->Q+Q-}).
In (\ref{rf1'}) we choose the first space to be the local quantum space
$\U_{-\ell}$ in site $k$, the second space to be the auxiliary
space $\U_{-s}\sim \C[z_0]$ and the third space to be
another copy of the auxiliary space $\U_{-s}\sim
\C[z_{0^{\prime}}]$
$$
\R^1_{00^{\prime}}(v_1|w_1,w_2) \,
\R_{k0^{\prime}}(u_{1},u_2|w_1,w_2) \,
\R_{k0}(u_{1},u_2|v_1,v_{2}) =
$$
\be \label{appPR1PRPR}
=\R_{k0}(u_{1},u_2|v_1,w_{2}) \,
\R_{k0^{\prime}}(u_1,u_{2}|w_1,v_{2}) \,
\R^1_{00^{\prime}}(v_1|w_1,w_2)
\ee
We have to
consider the appropriate limiting procedure.  We multiply this
relation by the dilatation operator $w_1^{z_0 \dd_0 + z_k \dd_k}$ from
the left and use \be \label{commut}
\left[z_0\partial_0+z_k\partial_k,\mathrm{R}_{k0}\right]=0 \,
\ee
to obtain\footnote{Here we underline
parameters to be excluded by taking asymptotics.}
$$
w_1^{z_0\partial_0} \, \R^1_{00^{\prime}}(v_1|\underline{w_1},w_2) \cdot
w_1^{z_k\partial_k} \, \R_{k0^{\prime}}(u_{1},u_2|\underline{w_1},w_2)
\cdot \R_{k0}(u_{1},u_2|v_1,v_{2}) =
$$
$$
=\R_{k0}(u_{1},u_2|v_1,w_{2}) \cdot w_1^{z_k\partial_k} \,
\R_{k0^{\prime}}(u_1,u_{2}|\underline{w_1},v_{2}) \cdot
w_1^{z_0\partial_0} \,
\R^1_{00^{\prime}}(v_1|\underline{w_1},w_2)
$$
Then we take limit $w_1 \to \infty$
with the help of (\ref{r+}), (\ref{R+})
$$
\mathbb{r}^{+}_{00^{\prime}}(v_1|\underline{w_2})\cdot
\R_{k0^{\prime}}^{+}(u_1,u_{2}|\underline{w_{2}})\cdot
\R_{k0}(u_{1},u_2|v_1,v_{2}) =
$$
$$
=\R_{k0}(u_{1},u_2|v_1,\underline{w_{2}})\cdot
\R_{k0^{\prime}}^{+}(u_1,u_{2}|v_{2})\cdot
\mathbb{r}^{+}_{00^{\prime}}(v_1|\underline{w_2})
$$
At the next step we multiply by $w_2^{-z_0 \dd_0 - z_k \dd_k}$ from the left
and use (\ref{commut}) again
$$
w_2^{-z_0 \dd_0} \,
\mathbb{r}^{+}_{00^{\prime}} (v_1|\underline{w_2}) \cdot
w_2^{-z_k \dd_k } \, \R_{k0^{\prime}}^{+}(u_1,u_{2}|\underline{w_{2}}) \cdot
\R_{k0}(u_{1},u_2|v_1,v_{2}) =
$$
$$
=\R_{k0}(u_{1},u_2|v_1,\underline{w_{2}}) w_2^{-z_k \dd_k} \cdot
\R_{k0^{\prime}}^{+}(u_1,u_{2}|v_{2}) \cdot
w_2^{-z_0 \dd_0 } \,
\mathbb{r}^{+}_{00^{\prime}}(v_1|\underline{w_2})
$$
Then we take the limit $w_2 \to \infty$ exploiting (\ref{2r+}), (\ref{v2R+}),
(\ref{R-}) and finally obtain (\ref{localT->-+})
$$
\mathrm{P}_{00^{\prime}} \,
(-)^{z_{0^{\prime}}\partial_{0^{\prime}}}\,
e^{z_0\partial_{0^{\prime}}} \cdot
e^{z_{0^{\prime}} \dd_k }
\cdot \R_{k0}(u_{1},u_2|v_1,v_{2}) =
$$
$$
= \R^{-}_{k0}(u_{1},u_2|v_1)\cdot
\R_{k0^{\prime}}^{+}(u_1,u_{2}|v_{2})\cdot
\mathrm{P}_{00^{\prime}} \,
(-)^{z_{0^{\prime}}\partial_{0^{\prime}}} \,
e^{z_0\partial_{0^{\prime}}}
$$
Similarly form (\ref{rf2'}) one can obtain the local factorization relation
$$
\mathbb{r}^{-}_{00^{\prime}}(v_{1}|v_2) \cdot
\mathrm{P}_{k0^{\prime}}\,\mathrm{r}^{+}_{k0^{\prime}}(u_1|u_2)
\,\mathrm{r}^{-}_{k0^{\prime}}(u_1|u_2)
\cdot \R_{k0}(u_{1},u_2|v_1,v_{2}) =
$$
\be \label{localT->+-}
=\R_{k0}^{+}(u_{1},u_2|v_{2}) \cdot
\R_{k0^{\prime}}^{-}(u_1,u_{2}|v_1)\cdot
\mathbb{r}^{-}_{00^{\prime}}(v_{1}|v_2)
\ee
which implies the first factorization in (\ref{T->Q+Q-}) if we take into account that
$$
\mathrm{P}_{12} \, \mathrm{r}^{+}(u_1|u_2) \, \mathrm{r}^{-} (u_1|u_2)=
\Gamma(z_{1}\partial_1+u_1-u_2+1) \,
e^{z_1\partial_2} \, \Gamma^{-1}(z_{1}\partial_1+u_1-u_2+1)
$$

Then we derive the local factorization relation underlying (\ref{Q-->Q1q-}).
In the relation (\ref{appPR1PRPR}) we choose $w_2 = u_2\,, \, v_1 = u_1$
$$
\R^1_{00^{\prime}}(v_1|w_1,u_2) \,
\R^{1}_{k0^{\prime}}(u_{1}|w_1,u_2) \,
\R^{2}_{k0}(u_{1},u_2|\underline{v_{2}}) =
$$
$$
=\mathrm{P}_{k0} \,
\R_{k0^{\prime}}(u_1,u_{2}|w_1,\underline{v_{2}}) \,
\R^1_{00^{\prime}}(u_1|w_1,u_2)\,
$$
then multiply by the dilatation operator $v_2^{-z_k \dd_k}$ from the right,
take the limit $v_2 \to \infty$ using (\ref{r-}) and
obtain the needed relation
(\ref{R-})
$$
\R^1_{00^{\prime}}(v_1|w_1,u_2) \cdot
\R^{1}_{k0^{\prime}}(u_{1}|w_1,u_2) \cdot
\mathbb{r}_{k0}^{-}(u_{1}|u_2) =
$$
$$
=\mathrm{P}_{k0} \cdot \R^{-}_{k0^{\prime}}(u_1,u_{2}|w_1) \cdot
\R^1_{00^{\prime}}(u_1|w_1,u_2)\,.
$$
Similarly (\ref{rf2'}) results in
$$
\R^2_{00^{\prime}}(u_1,v_2|u_2) \cdot
\mathbb{r}^{+}_{k0'}(u_{1}|u_2) \cdot
\R^{2}_{k0}(u_{1},u_2|v_{2}) =
$$
$$
=\R^{-}_{k0}(u_{1},u_2|v_{2}) \cdot
\mathrm{P}_{k0^{\prime}} \cdot
\R^2_{00^{\prime}}(u_{1},v_{2}|u_2)
$$
implying the factorization (\ref{Q+->q+Q2}).

\subsection{Local commutativity}

Now we are going to present the local relations which underlie
commutativity relations for transfer matrices. For this purpose we use
the general Yang-Baxter relation
(\ref{rf'}). Rewriting it in the form
\be
\label{PRPRPR} \R_{00^{\prime}}(v_{1},v_2|w_{1},w_2)
\R_{k0^{\prime}}(u_{1},u_2|w_1,w_2) \R_{k0}(u_1,u_2|v_{1},v_2)=
$$
$$
=\R_{k0}(u_{1},u_{2}|v_1,v_2)
\R_{k0^{\prime}}(u_1,u_{2}|w_1,w_{2})
\R_{00^{\prime}}(v_{1},v_2|w_{1},w_2)
\ee
we derive  immediately the commutativity of
the general transfer matrices (\ref{T}).
The commutativity for the other transfer
matrices follows from
$$
\R_{00^{\prime}}(u_{1},v_2|w_{1},u_2)
\R^{1}_{k0^{\prime}}(u_{1}|w_1,u_2) \R^{2}_{k0}(u_1,u_2|v_2)=
$$
\be \label{RR1R2} =\R^{2}_{k0}(u_{1},u_{2}|v_2)
\R^{1}_{k0^{\prime}}(u_1|w_1,u_{2})
\R_{00^{\prime}}(u_{1},v_2|w_{1},u_2)
\ee

$$
\R^{1}_{00^{\prime}}(v_{1}|w_{1},u_2)
\R^{1}_{k0^{\prime}}(u_{1}|w_1,u_2) \R^{1}_{k0}(u_1|v_1,u_2)=
$$
\be \label{R1R1R1} =\R^{1}_{k0}(u_{1}|v_{1},u_2)
\R^{1}_{k0^{\prime}}(u_1|w_1,u_{2})
\R^{1}_{00^{\prime}}(v_{1}|w_{1},u_2)
\ee

$$
\R^{2}_{00^{\prime}}(u_{1},v_2|w_{2})
\R^{2}_{k0^{\prime}}(u_{1},u_2|w_2) \R^{2}_{k0}(u_1,u_2|v_2) =
$$
\be \label{R2R2R2} =\R^{2}_{k0}(u_{1},u_2|v_2)
\R^{2}_{k0^{\prime}}(u_1,u_2|w_{2})
\R^{2}_{00^{\prime}}(u_{1},v_2|w_2)
\ee
which are obtained from (\ref{PRPRPR})
specifying parameters  (for more details see part I, Appendix A).

We start with (\ref{R1R1R1})
multiply it by $w_1^{z_0 \dd_0 + z_k \dd_k}$ from the left
$$
w_1^{z_0 \dd_0} \, \R^{1}_{00^{\prime}}(v_{1}|\underline{w_{1}},u_2)
\cdot w_1^{z_k \dd_k}  \,
\R^{1}_{k0^{\prime}}(u_{1}|\underline{w_1},u_2) \cdot
\R^{1}_{k0}(u_1|v_1,u_2)=
$$
$$
=\R^{1}_{k0}(u_{1}|v_{1},u_2) \cdot w_1^{z_k \dd_k}
\,\R^{1}_{k0^{\prime}}(u_1|\underline{w_1},u_{2}) \cdot w_1^{z_0
\dd_0} \, \R^{1}_{00^{\prime}}(v_{1}|\underline{w_{1}},u_2)
$$
and do the limit $w_1 \to \infty$ by using (\ref{r+})
$$
\mathbb{r}^{+}_{00^{\prime}}(v_{1}|u_2) \cdot
\mathbb{r}^{+}_{k0^{\prime}}(u_{1}|u_2) \cdot
\R^{1}_{k0}(u_1|v_1,u_2)=
$$
\be \label{intertwiner+R1} =\R^{1}_{k0}(u_{1}|v_{1},u_2) \cdot
\mathbb{r}^{+}_{k0^{\prime}}(u_{1}|u_2) \cdot
\mathbb{r}^{+}_{00^{\prime}}(v_{1}|u_2) \,.
\ee
The last relation implies the commutativity of the
transfer matrices constructed from $\mathbb{r}^{+}$ and
$\R^{1}$. In much the same way (\ref{R2R2R2}) produces
$$
\mathbb{r}^{-}_{00^{\prime}}(u_{1}|v_2) \cdot
\mathbb{r}^{-}_{k0^{\prime}}(u_{1}|u_2) \cdot
\R^{2}_{k0}(u_1,u_2|v_2)=
$$
\be \label{intertwiner-R2} =\R^{2}_{k0}(u_{1},u_{2}|v_2) \cdot
\mathbb{r}^{-}_{k0^{\prime}}(u_{1}|u_2) \cdot
\mathbb{r}^{-}_{00^{\prime}}(u_{1}|v_2)\,
\ee
resulting in the commutativity of the transfer-matrices built from
$\mathbb{r}^{-}$ and $\R^{2}$.

In order to obtain the other commutation relations we use (\ref{RR1R2})
multiply it by $w_1^{z_0 \dd_0 + z_k \dd_k}$ from the left
$$
w_1^{z_0 \dd_0} \, \R_{00^{\prime}}(u_{1},v_2|\underline{w_{1}},u_2)
\cdot w_1^{z_k \dd_k} \,
\R^{1}_{k0^{\prime}}(u_{1}|\underline{w_1},u_2) \cdot
\R^{2}_{k0}(u_1,u_2|v_2)=
$$
$$
=\R^{2}_{k0}(u_{1},u_{2}|v_2) \cdot w_1^{z_k \dd_k} \,
\R^{1}_{k0^{\prime}}(u_1|\underline{w_1},u_{2}) \cdot w_1^{z_0
\dd_0} \, \R_{00^{\prime}}(u_{1},v_2|\underline{w_{1}},u_2)
$$
and take the limit $w_1 \to \infty$ by means of (\ref{r+}), (\ref{R+})
$$
\R^{+}_{00^{\prime}}(u_{1},v_2|u_2) \cdot
\mathbb{r}^{+}_{k0^{\prime}}(u_{1}|u_2) \cdot
\R^{2}_{k0}(u_1,u_2|v_2)=
$$
\be \label{intertwiner+R2} =\R^{2}_{k0}(u_{1},u_{2}|v_2) \cdot
\mathbb{r}^{+}_{k0^{\prime}}(u_{1}|u_2) \cdot
\R^{+}_{00^{\prime}}(u_{1},v_2|u_2)\,. \ee
This means that the
transfer matrices constructed from $\mathbb{r}^{+}$ and
$\R^{2}$ commute. Similarly it is easy to derive the
intertwining relation for $\mathbb{r}^{-}$ and $\R^{1}$
$$
\R^{-}_{00^{\prime}}(v_{1},u_2|u_{1}) \cdot
\mathbb{r}^{-}_{k0^{\prime}}(u_{1}|u_2) \cdot
\R^{1}_{k0}(u_1|v_1,u_2)=
$$
\be \label{intertwiner-R1} =\R^{1}_{k0}(u_{1}|v_{1},u_2) \cdot
\mathbb{r}^{-}_{k0^{\prime}}(u_{1}|u_2) \cdot
\R^{-}_{00^{\prime}}(v_{1},u_2|u_{1})\,. \ee

In principle combining commutativity relations (produced by local relations presented above)
with factorization relations for transfer matrices
we deduce immediately the commutativity for the all transfer
matrices. However it is also possible to
prove the commutativity relations from
the local intertwining relations
$$
\R^{+}_{00^{\prime}}(v_{1},v_2|w_2) \cdot
\R^{+}_{k0^{\prime}}(u_{1},u_2|w_2) \cdot
\R_{k0}(u_1,u_2|v_{1},v_2)=
$$
\be \label{intertwineR+R} = \R_{k0}(u_{1},u_{2}|v_1,v_2) \cdot
\R^{+}_{k0^{\prime}}(u_1,u_{2}|w_{2}) \cdot
\R^{+}_{00^{\prime}}(v_{1},v_2|w_2)\,.
\ee

$$
\R^{++}_{00^{\prime}}(v_2|w_2) \cdot
\R^{+}_{k0^{\prime}}(u_{1},u_2|w_2) \cdot \R^{+}_{k0}(u_1,u_2|v_2)=
$$
\be \label{intertwineR+R+} =\R^{+}_{k0}(u_{1},u_{2}|v_2) \cdot
\R^{+}_{k0^{\prime}}(u_1,u_{2}|w_{2}) \cdot
\R^{++}_{00^{\prime}}(v_2|w_2) \,
\ee

$$
\R^{-}_{00^{\prime}}(v_{1},v_2|w_{1}) \cdot
\R^{-}_{k0^{\prime}}(u_{1},u_2|w_1) \cdot
\R_{k0}(u_1,u_2|v_{1},v_2)=
$$
\be \label{intertwineR-R} =\R_{k0}(u_{1},u_{2}|v_1,v_2) \cdot
\R^{-}_{k0^{\prime}}(u_1,u_{2}|w_1) \cdot
\R^{-}_{00^{\prime}}(v_{1},v_2|w_{1})\,.
\ee

$$
\R^{--}_{00^{\prime}}(v_{1}|w_{1}) \cdot
\R^{-}_{k0^{\prime}}(u_{1},u_2|w_1) \cdot
\R^{-}_{k0}(u_1,u_2|v_{1})=
$$
\be \label{intertwineR-R-} =\R^{-}_{k0}(u_{1},u_{2}|v_1) \cdot
\R^{-}_{k0^{\prime}}(u_1,u_{2}|w_1) \cdot
\R^{--}_{00^{\prime}}(v_{1}|w_{1}) \,.
\ee
which can be obtained from (\ref{PRPRPR}) performing
appropriate limiting procedures in much the same way as before.

\section{Finite-dimensional construction}

\setcounter{equation}{0}

Here we perform some calculations which concern the Baxter operator construction
for finite-dimensional representations relying on part I, section 5.

\subsection{Restriction of the general $\R$-operator to finite-dimensional
representations}

In part I, subsection 5.2, we have obtained as the explicit form
of the restricted $\R$-operator
$$
\mathbf{R}_{12}\left(u|{\textstyle\frac{n}{2}}, s\right) =
\mathrm{P}_{12}\cdot e^{-z_1 \dd_2}\frac{(-1)^{z_2\partial_2}\,
}{\Gamma(z_2\partial_2 -\frac{n}{2}-s-u)\Gamma (1+n
-z_2\partial_2)} \, \Pi^n_2 \, e^{z_1 \dd_2} \cdot
$$
\be \label{Norm1}
\cdot e^{-z_2 \dd_1}\, (-1)^{z_1\partial_1}\Gamma(z_1\partial_1
-\textstyle{\frac{n}{2}}-s+u)\,\Gamma (1+n -z_1\partial_1)\,
e^{z_2 \dd_1} \Pi_1^n\,.
\ee
Recall also that finite-dimensional subspace is invariant
under the action of operator
$\mathbf{R}_{12}\left(u|{\textstyle\frac{n}{2}}, s \right)$
because projector $\Pi_1^n$ appears not only on the right but on
the left too
\be \label{Sproperty}
\mathrm{P}_{12}\, e^{-z_1\partial_2}\, \Pi_2^n = \mathrm{P}_{12}\,
\Pi_2^n \, e^{-z_1\partial_2}\, \Pi_2^n = \Pi_1^n\,
\mathrm{P}_{12}\,e^{-z_1\partial_2}\, \Pi_2^n\,.
\ee

In the parametrization (\ref{paramfindim})
$$
u_1 = u - {\textstyle\frac{n}{2}} -1\ , \ u_2 = u +
{\textstyle\frac{n}{2}}\quad ;\quad v_1 = v - s -1\ , \ v_2 = v + s\
$$
we rewrite (\ref{Norm1}) as follows
\begin{equation}\label{Norm2}
\mathbf{R}_{12}(u_1 , u_2|v_1,v_2)
= \mathrm{P}_{12}\cdot e^{-z_1
\dd_2}\,\frac{(-1)^{z_2\partial_2}}{\Gamma(z_2\partial_2
+v_1-u_2+1)\Gamma (u_2-u_1-z_2\partial_2)}\, \Pi^n_2\, e^{z_1
\dd_2}\cdot
\end{equation}
$$
\cdot e^{-z_2 \dd_1}\, (-1)^{z_1\partial_1}\, \Gamma(z_1\partial_1
+u_1-v_2+1)\Gamma (u_2-u_1-z_1\partial_1)\, e^{z_2 \dd_1}\,
\Pi_1^n\,.
$$

Doing  similar calculations as with the operators $\R^{+}$ and $\R^{-}$
considered in subsection 3.2 we obtain

\begin{itemize}

\item Operator $\mathbf{R}^{+}$
$$
\mathbf{R}^{+}_{12}\left(u_1,u_2|v_2\right) =
\mathrm{P}_{12}\cdot \frac{(-1)^{z_2\partial_2}\,
}{\Gamma (u_2-u_1-z_2\partial_2)} \, \Pi^n_2 \, e^{z_1 \dd_2} \cdot
$$
\be \label{bfR+Expl}
\cdot e^{-z_2 \dd_1}\, (-1)^{z_1\partial_1}\Gamma(z_1\partial_1+
u_1-v_2+1)\,\Gamma (u_2-u_1 -z_1\partial_1)\,
e^{z_2 \dd_1} \Pi_1^n\,
\ee

\item Operator $\mathbf{R}^{-}$
$$
\mathbf{R}^{-}_{12}\left(u_1,u_2|v_1\right) =
\mathrm{P}_{12}\cdot e^{-z_1 \dd_2}\frac{(-1)^{z_2\partial_2}\,
}{\Gamma(z_2\partial_2 + v_1 - u_2 + 1)\Gamma(u_2-u_1-z_2\partial_2)}
\, \Pi^n_2 \, e^{z_1 \dd_2} \cdot
$$
\be \label{bfR-Expl}
\cdot e^{-z_2 \dd_1}\, (-1)^{z_1\partial_1}\,\Gamma (u_2-u_1 -z_1\partial_1)\, \Pi_1^n\,
\ee

One can easily see that $\mathbf{R}^{-}$ and $\mathbf{R}^{+}$
do not map beyond the subspace $\V_n\otimes\U_{-s}$
as well as $\mathbf{R}$. The same is true for
$\mathbf{R}^1,\,\mathbf{R}^2,\,\mathbf{r}^{+}$ and $\mathbf{r}^{-}$.
In all these operators the parameters are as in (\ref{paramfindim}).

Previously we have calculated the double reduction
of $\mathrm{R}$-operator (\ref{bfv2R+}). The finite-dimensional
analogue of this formula is
\be \label{bfv2R+}
v_2^{-z_1 \dd_1} \, \mathbf{R}^{+}(u_1,u_2|v_{2})
\to \mathrm{P}_{12} \, e^{z_2 \dd_1}\, \Pi^n_1
\ee

\item Operator $\mathbf{R}^1$

Explicit expressions for $\mathbf{R}^1,\,\mathbf{R}^2$ are obtained
in part I:
\begin{equation}\label{R1findim}
\mathbf{R}^1_{12}(u_1 |v_1,u_2) \equiv \mathrm{P}_{12}\cdot e^{-z_1
\dd_2} \frac{(-1)^{z_2\partial_2+u_2-u_1-1}}{\Gamma(z_2\partial_2
+v_1-u_2+1)\Gamma (u_2-u_1-z_2\partial_2)}\,\Pi^n_2\, e^{z_1
\dd_2} \cdot \Pi_1^n
\end{equation}

\item Operator $\mathbf{R}^{2}$
\begin{equation}\label{R2findim}
\mathbf{R}^2_{12}(u_1,u_2 |v_2)  \equiv \mathrm{P}_{12}\cdot e^{-z_1
\dd_2}\, \Pi^n_2\, e^{z_1 \dd_2}\cdot
\end{equation}
$$
\cdot e^{-z_2 \dd_1}\,
(-1)^{z_1\partial_1+u_2-u_1-1}\,\Gamma(z_1\partial_1 +u_1-v_2+1)\,
\Gamma (u_2-u_1-z_1\partial_1)\, e^{z_2 \dd_1}\, \Pi_1^n
$$
Comparing expressions of $\mathbf{R}^1$ with $\mathbf{R}^{-}$ and
of $\mathbf{R}^2$ with $\mathbf{R}^{+}$ we that they are very
similar. Indeed they  are all derived from the same
$\mathbf{R}$-operator.

\item Operator $\mathbf{r}^{+}$

In order to calculate $\mathbf{r}^{+},\,\mathbf{r}^{-}$ we start
from definitions (\ref{bfr+}), (\ref{bfr-}) and use Stirling's
formula (\ref{Stirling}) \be \label{APPbfr+}
\mathbf{r}^{+}_{12}(u_1|u_2) = \mathrm{P}_{12}\cdot
\frac{(-1)^{z_2\partial_2 + u_2-u_1-1}\, }{\Gamma
(u_2-u_1-z_2\partial_2)} \, \Pi^n_2 \, e^{z_1 \dd_2} \cdot \Pi_1^n
\ee

\item Operator $\mathbf{r}^{-}$
 \be \label{APPbfr-}
\mathbf{r}^{-}_{12}(u_1|u_2) = \mathrm{P}_{12}\cdot
e^{-z_1\dd_2}\, \Pi^n_2\, e^{z_1 \dd_2}\cdot e^{-z_2 \dd_1}\,
(-1)^{u_2-u_1-1}\,\Gamma (u_2-u_1 -z_1\partial_1)\, \Pi_1^n \ee

\end{itemize}

\subsection{Local factorization}

The derivation of the local relations in the case
of finite-dimensional operators goes parallel to the calculation
in the infinite-dimensional case.
Let us obtain (\ref{BFlocalT->-+}).
In (\ref{rf1'}) we choose the first space to be the local quantum space
$\U_{-\ell}$ in $k$-site, the second space to be the  auxiliary
space $\U_{-s}\sim \C[z_0]$ and the third space to be
another auxiliary space $\U_{-s}\sim
\C[z_{0^{\prime}}]$ and then restrict   the
quantum space to the finite-dimensional subspace $\V_{n}$
at $\ell=\frac{n}{2}$
$$
\R^1_{00^{\prime}}(v_1|w_1,w_2) \,
\mathbf{R}_{k0^{\prime}}(u_{1},u_2|w_1,w_2) \,
\mathbf{R}_{k0}(u_{1},u_2|v_1,v_{2}) =
$$
$$
=\mathbf{R}_{k0}(u_{1},u_2|v_1,w_{2}) \,
\mathbf{R}_{k0^{\prime}}(u_1,u_{2}|w_1,v_{2}) \,
\R^1_{00^{\prime}}(v_1|w_1,w_2)
$$
We have to
consider the appropriate limiting procedure.  We multiply this
relation by the dilatation operator $w_1^{z_0 \dd_0 + z_k \dd_k}$ from
the left
to obtain
$$
w_1^{z_0\partial_0} \, \R^1_{00^{\prime}}(v_1|\underline{w_1},w_2) \cdot
w_1^{z_k\partial_k} \, \mathbf{R}_{k0^{\prime}}(u_{1},u_2|\underline{w_1},w_2)
\cdot \mathbf{R}_{k0}(u_{1},u_2|v_1,v_{2}) =
$$
$$
=\mathbf{R}_{k0}(u_{1},u_2|v_1,w_{2}) \cdot w_1^{z_k\partial_k} \,
\mathbf{R}_{k0^{\prime}}(u_1,u_{2}|\underline{w_1},v_{2}) \cdot
w_1^{z_0\partial_0} \,
\R^1_{00^{\prime}}(v_1|\underline{w_1},w_2)
$$
Then we take the limit $w_1 \to \infty$
with the help of (\ref{r+}), (\ref{bfR+})
$$
\mathbb{r}^{+}_{00^{\prime}}(v_1|\underline{w_2})\cdot
\mathbf{R}_{k0^{\prime}}^{+}(u_1,u_{2}|\underline{w_{2}})\cdot
\mathbf{R}_{k0}(u_{1},u_2|v_1,v_{2}) =
$$
$$
=\mathbf{R}_{k0}(u_{1},u_2|v_1,\underline{w_{2}})\cdot
\mathbf{R}_{k0^{\prime}}^{+}(u_1,u_{2}|v_{2})\cdot
\mathbb{r}^{+}_{00^{\prime}}(v_1|\underline{w_2})
$$
At the next step we multiply by $w_2^{-z_0 \dd_0 - z_k \dd_k}$ from the left
$$
w_2^{-z_0 \dd_0} \,
\mathbb{r}^{+}_{00^{\prime}} (v_1|\underline{w_2}) \cdot
w_2^{-z_k \dd_k } \, \mathbf{R}_{k0^{\prime}}^{+}(u_1,u_{2}|\underline{w_{2}}) \cdot
\mathbf{R}_{k0}(u_{1},u_2|v_1,v_{2}) =
$$
$$
=\mathbf{R}_{k0}(u_{1},u_2|v_1,\underline{w_{2}}) w_2^{-z_k \dd_k} \cdot
\mathbf{R}_{k0^{\prime}}^{+}(u_1,u_{2}|v_{2}) \cdot
w_2^{-z_0 \dd_0 } \,
\mathbb{r}^{+}_{00^{\prime}}(v_1|\underline{w_2})
$$
Then we take the limit $w_2 \to \infty$ using (\ref{2r+}), (\ref{bfv2R+}),
(\ref{bfR-}) and finally obtain (\ref{BFlocalT->-+})
$$
\mathrm{P}_{00^{\prime}} \,
(-)^{z_{0^{\prime}}\partial_{0^{\prime}}}\,
e^{z_0\partial_{0^{\prime}}} \cdot
e^{z_{0^{\prime}} \dd_k }\,\Pi^n_k
\cdot \mathbf{R}_{k0}(u_{1},u_2|v_1,v_{2}) =
$$
$$
= \mathbf{R}^{-}_{k0}(u_{1},u_2|v_1)\cdot
\mathbf{R}_{k0^{\prime}}^{+}(u_1,u_{2}|v_{2})\cdot
\mathrm{P}_{00^{\prime}} \,
(-)^{z_{0^{\prime}}\partial_{0^{\prime}}} \,
e^{z_0\partial_{0^{\prime}}}
$$

\subsection{Local commutativity}

In order to illustrate the derivation of the local commutativity relation
we are going to derive the  underlying commutativity of the transfer matrices
$\mathbf{r}^{+}$ and $\mathbf{R}^{1}$. We start with
the general Yang-Baxter equation restricted
to the space $\V^n\otimes \C[z_0]\otimes \C[z_{0'}]$
$$
\R_{00^{\prime}}(v_{1},v_2|w_{1},w_2)
\mathbf{R}_{k0^{\prime}}(u_{1},u_2|w_1,w_2) \mathbf{R}_{k0}(u_1,u_2|v_{1},v_2)=
$$
$$
=\mathbf{R}_{k0}(u_{1},u_{2}|v_1,v_2)
\mathbf{R}_{k0^{\prime}}(u_1,u_{2}|w_1,w_{2})
\R_{00^{\prime}}(v_{1},v_2|w_{1},w_2)
$$
choose parameters $w_2 = u_2 - \delta,\,v_2 = u_2 - \delta$
and take the limit $\delta \to 0$ (\ref{bfR1})
$$
\R^{1}_{00^{\prime}}(v_{1}|w_{1},u_2)
\mathbf{R}^{1}_{k0^{\prime}}(u_{1}|w_1,u_2) \mathbf{R}^{1}_{k0}(u_1|v_1,u_2)=
$$
\be  =\mathbf{R}^{1}_{k0}(u_{1}|v_{1},u_2)
\mathbf{R}^{1}_{k0^{\prime}}(u_1|w_1,u_{2})
\R^{1}_{00^{\prime}}(v_{1}|w_{1},u_2)
\ee
Then we multiply it by $w_1^{z_0 \dd_0 + z_k \dd_k}$ from the left
$$
w_1^{z_0 \dd_0} \, \R^{1}_{00^{\prime}}(v_{1}|\underline{w_{1}},u_2)
\cdot w_1^{z_k \dd_k}  \,
\mathbf{R}^{1}_{k0^{\prime}}(u_{1}|\underline{w_1},u_2) \cdot
\mathbf{R}^{1}_{k0}(u_1|v_1,u_2)=
$$
$$
=\mathbf{R}^{1}_{k0}(u_{1}|v_{1},u_2) \cdot w_1^{z_k \dd_k}
\,\mathbf{R}^{1}_{k0^{\prime}}(u_1|\underline{w_1},u_{2}) \cdot w_1^{z_0
\dd_0} \, \R^{1}_{00^{\prime}}(v_{1}|\underline{w_{1}},u_2)
$$
and do the limit $w_1 \to \infty$ using (\ref{r+}), (\ref{bfr+})
$$
\mathbb{r}^{+}_{00^{\prime}}(v_{1}|u_2) \cdot
\mathbf{r}^{+}_{k0^{\prime}}(u_{1}|u_2) \cdot
\mathbf{R}^{1}_{k0}(u_1|v_1,u_2)=
$$
$$ =\mathbf{R}^{1}_{k0}(u_{1}|v_{1},u_2) \cdot
\mathbf{r}^{+}_{k0^{\prime}}(u_{1}|u_2) \cdot
\mathbb{r}^{+}_{00^{\prime}}(v_{1}|u_2) \,.
$$

\end{document}